\documentstyle[12pt,aasms4]{article}

\lefthead{Pereyra, Kallman, \& Blondin}
\righthead{Line-Driven Accretion Disk Winds II: Nonisothermal Model}

\begin{document}

\title{Hydrodynamic Models of Line-Driven Accretion Disk Winds II:
       Adiabatic Winds from Nonisothermal Disks}

\author{Nicolas Antonio Pereyra\altaffilmark{1} and Timothy R. Kallman}
\affil{NASA/GSFC, Laboratory for High Energy Astrophysics,
       Code 662, Greenbelt, MD 20771; \
       pereyra@horus.gsfc.nasa.gov, tim@xstar.gsfc.nasa.gov}

\and

\author{John M. Blondin}
\affil{North Carolina State University, Department of Physics, Box 8202,
       Raleigh, NC 27695; \
       blondin@eeyore.physics.ncsu.edu}

\altaffiltext{1}{Department of Astronomy,
                 University of Maryland, College Park, MD 20742-2421}

\begin{abstract}
We present here numerical hydrodynamic simulations of line-driven
accretion disk winds in cataclysmic variable systems.
We calculate wind mass-loss rate,
terminal velocities,
and line profiles for \ion{C}{4} (1550~\AA) for various viewing angles.
The models are 2.5-dimensional,
include an energy balance condition,
and calculate the radiation field as a function of position near an
optically thick accretion disk.
The model results show that centrifugal forces produce collisions of
streamlines in the disk wind which in turn generate an enhanced
density region,
underlining the necessity of two dimensional calculations where these
forces may be represented.
For disk luminosity $L_{disk}=L_{\sun}$,
white dwarf mass $M_{wd}=0.6M_{\sun}$,
and white dwarf radii $R_{wd}=0.01R_{\sun}$,
we obtain a wind mass-loss rate of
$\dot M_{wind}=8 \times 10^{-12} M_{\sun} {\rm yr}^{-1}$,
and a terminal velocity of $\sim 3000 {\rm \; km \; s}^{-1}$.
The line profiles we obtain are consistent with observations in their
general form,
in particular in the maximum absorption at roughly half the terminal
velocity for the blue-shifted component,
in the magnitudes of the wind velocities implied by the absorption
components,
in the FWHM of the emission components,
and in the strong dependence in inclination angle.
\end{abstract}


\section{Introduction}

The first evidence for winds from cataclysmic variables (CVs) came
from the discovery of P~Cygni profiles in the UV resonance lines of
SS~Cygni
(\cite{hea85}).
The dependence on the observability of CV winds with inclination
angle and the apparent similarity between cataclysmic variable
line profiles and those of OB stars led to the early suggestion
by C\'ordova \& Mason (1982) that the winds in CVs originate from the
accretion disk and that the line radiation pressure is responsible for
the wind.
More recently,
P~Cygni profiles have been detected from virtually all nonmagnetic CVs
with high mass accretion rates
($\gtrsim 4\times 10^{-10} M_{\sun} {\rm yr}^{-1}$)
(e.g., \cite{cor82}; \cite{gre82}; \cite{gui82}; \cite{pri95};
       \cite{fri97}; \cite{gan97}; \cite{kni97}).
The absorption component is most apparent in low inclination systems,
and is not detected in eclipsing systems
(e.g.,
\cite{hut80}; \cite{kra81}; \cite{hol82}; 
\cite{cor85}; \cite{mas95}; \cite{che97}).

Past efforts in the development of models for CV winds focused on
one-dimensional models (\cite{vit88}; \cite{kal88}), and on kinematic
modeling (\cite{shl93}; \cite{kni95}), which succeeded in showing
consistency between the assumed polar geometry of a disk wind and
observed profiles.

Icke~(1980) developed a two-dimensional disk wind model but did not
take into account the radiation pressure due to line absorption.
For a typical white dwarf mass of $0.6M_{\sun}$ (\cite{lei80})
the disk luminosity required to produce such a wind without line
radiation pressure
(assuming radiation pressure due to continuum scattering only)
would require a luminosity of $\sim 10^5 L_{\sun}$ (\cite{ick80}),
several magnitudes above the observed luminosity of CVs which vary
between $0.01L_{\sun}$ and $10 L_{\sun}$ (\cite{pat84}).
Icke did find that for accretion disk winds in general,
for sufficiently high disk luminosity,
a biconical disk wind would be produced.
Icke~(1981) suggested that biconical winds were a general property
of accretion disk winds independent of the wind driving mechanism.

Murray et~al. (1995) found that for line-driven disk winds in
active galactic nuclei (AGN) the wind flow tends to be parallel to the
accretion disk and developed a one-dimensional disk wind model for
these systems.
Murray \& Chiang (1996) found that the
single-peaked optical emission lines seen in AGNs and 
high-luminosity CVs could be accounted
for by emission at the base of an accretion disk wind.
Using the disk wind model of Murray et~al. (1995),
Murray \& Chiang (1997) calculated synthetic line profiles
of \ion{C}{4} 1550\AA \ and found qualitative agreement 
between their models and observed line profiles.
In the case of CVs,
as we discussed above,
the winds are most apparent in low-inclination systems and not
observed in eclipsing systems,
indicating that the winds in CVs tend to flow perpendicular to the
accretion disk.

In an earlier paper (\cite{per97}, hereafter Paper~I) we presented
two-dimensional isothermal hydrodynamic models of line-driven
accretion disk winds (LDADW) in CVs.
To our knowledge these isothermal models were the first
two-dimensional hydrodynamic models developed for LDADWs.
Results from Paper~I show,
in analogy with line-driven winds from early type stars,
that terminal velocities are approximately independent
of the luminosity of the disk, although increments in luminosity
produce increments in mass-loss rate. 
In Paper~I we showed that rotational forces are important in the study
of winds from accretion disks,
and that they cause the velocity streamlines to collide.
The collision of streamlines reduce the speed and increase the density
of the wind producing an enhanced density region.
In Paper~I we also showed that the highest absorption occurs in the
enhanced density region where density is increased relative to a
spherically diverging wind with the same mass loss rate and the
velocity is roughly half the terminal velocity.
This density increase is necessary in order to produce at least
marginally optically thick lines.

Recently Proga, Stone, \& Drew (1998) also developed two-dimensional
models for these systems,
obtaining similar results to the models presented here.
When evaluating the line-radiation pressure from the disk,
in order to make the computations manageable,
they assumed that the velocity gradient is primarily along
the ``z'' direction,
as we also did in Paper~1 and in the models presented here.
Proga, Stone, \& Drew worked out their hydrodynamic model in
spherical coordinates, while the models presented here were
implemented in cylindrical coordinates.
They found similar wind velocities and wind mass loss rates
as we do here,
although they assumed disk luminosities about an order
of magnitude greater.
A difference we find with the results of the models presented by
Proga, Stone \& Drew (1998) is that they obtain unsteady flows
characterized by large amplitude fluctuations in velocity and density,
while for the models presented in this paper we find a steady flow.

A shortcoming of our previous work,
presented in Paper~I,
was that we had assumed an isothermal wind and an isothermal disk.
In this work we develop a hydrodynamic model of LDADWs in
CVs which includes the radial structure of an optically thick accretion
disk with the corresponding radiation fields and surface temperature
distributions.
We have also implemented the corresponding
energy conservation equation for the models presented here
self consistently (see equation~[18]),
including the adiabatic heating and cooling effects due to
compression and expansion.

For the models presented in this paper we have also taken
values presented by Abbott~(1982) for the line radiation force
multiplier parameters,
rather than the earlier values used by Castor, Abbott, \& Klein~(1975)
as we did in our isothermal models presented in Paper~I. 

From the results of the hydrodynamic model presented in this paper,
we calculate theoretical line profiles for the 
\ion{C}{4} 1550~\AA \ line,
and find that the line profiles obtained through our model are
consistent with observations in their general form and strong
dependence with inclination angle.
Thus we are able to predict the general observed wind properties of CVs.

The boundary layer where the disk intercepts the white dwarf and the
white dwarf photosphere may also contribute to the radiation field.
The white dwarf by itself
(without considering the effects of the accreting mass)
will have a luminosity of the order of $0.01L_{\sun}$(\cite{lei80}),
while the accretion disk will have typical luminosities of the
order of $L_{\sun}$.
In a steady state accretion disk half of the energy of the accreting
mass is converting into radiation emitted by the accretion disk.
Assuming a relatively slowly rotating compact star,
it has been predicted that the boundary layer presents a luminosity
approximately equal to the accretion disk,
thus accounting for the other half of the energy of the accreting mass
(e.g., \cite{fra92}).
But there is still uncertainty concerning the existence and spectrum
of the boundary layer
(\cite{vrt94}; \cite{mau95}; \cite{lon96}).
Analysis of observations have found,
in some cases,
a boundary layer luminosity an order of magnitude less than the disk
luminosity (e.g., \cite{mau95}),
and in other cases a boundary layer luminosity comparable
to disk luminosity (\cite{lon96}).
In the models presented here we have assumed that the
radiation field is generated by the accretion disk alone and
have therefore neglected any boundary layer or white dwarf
radiation.
In the future we plan to explore the effects of boundary layer radiation
on our models.

In \S~2 we discuss the radial structure of the accretion disk
and the radiation field as implemented in our model.
In \S~3 we derive the expression we use for the treatment of the
line radiation pressure.
In \S~4 we present and discuss the hydrodynamic calculations. 
In \S~5 we present and discuss theoretical
\ion{C}{4} 1550~\AA \ line profiles and compare them with observations.
In \S~6 we present a summary and conclusions of this work.

\section{Accretion Disk}

Cataclysmic variables are binary systems composed of a white dwarf,
a main sequence star,
and an accretion disk about the white dwarf.
The disk is formed by mass accreting from the main sequence star
to the white dwarf.
In some objects strong magnetic fields
($\gtrsim 10^7 {\rm G}$)
may prevent the accreting mass in CVs from forming a disk around the
white dwarf
(e.g., \cite{fra92}).
In this work we study the winds originating in nonmagnetic CVs
in which an accretion disk is formed.

Direct observational evidence for the existence of accretion disks in
CVs is found in the spectrum of eclipsing CVs.
Eclipsing CVs present double peaked H and He emission lines in their
optical spectrum.
In these systems the eclipse of the blue side of the emission line
precedes that of the red side
(\cite{gre59}; \cite{you81}; \cite{mar90}),
demonstrating that the emission lines occupy an extended region and
that it rotates about white dwarf in the direction expected for a disk
formed through the accretion of mass from the secondary star.
Furthermore,
synthetic emission lines that have been calculated from models of
accretion disks rotating around the white dwarf have succeeded in
accounting for the double peak features
found in observations 
(\cite{sma69}, 1981; \cite{hua72}; \cite{hor86}).

We wish to note here that single peaked,
rather than double peaked,
H and He emission lines with the same eclipse behavior
(eclipse of the blue wing preceding eclipse of the red wing)
are observed in many CVs
(e.g., \cite{hon86}; \cite{tho91}; \cite{beu92}).
These single peaked Balmer emission lines may be explained
by the Stark effect in the disk radiated emission
(\cite{lin88}; \cite{fer97}) or by line emission at the base
of an accretion disk wind (\cite{mur96}).

Further evidence for the existence accretion disks has been obtained
through models for disk structure that have succeeded in accounting for
the observed continuum shapes in many CVs
(e.g., \cite{kip79}; \cite{fra81}; \cite{wad88}; \cite{hor94}).
In most cases the luminosity of CVs is dominated by the radiation
emitted by the accretion disk which is typically of order of
the solar luminosity $L_{\sun}$ (\cite{pat84}).
CVs with high mass accretion rates (dwarf novae in outburst,
novae, and nova-like variables) have luminosities which range
between $0.5 L_{\sun}$ and $10 L_{\sun}$ (\cite{war87}).

In cataclysmic variables the disk radiation emission and disk
temperature are radially dependent.
Such disks have been studied in detail by Shakura and Sunyaev~(1973)
and we adopt their formulation in what follows.
The accretion disk would have an inner radius approximately equal to
the radius of the white dwarf
(\cite{pri81}; \cite{pac91}; \cite{pop91}, 1995)
and an outer radius reaching out to two to three times less than
the inner Lagrangian point (\cite{dhi91}; \cite{sma94}; \cite{sti95})
(see Figures~1 and 2).
Assuming that the disk is in a steady state,
that the shear stress at the inner radius of the disk is negligible,
and that the gas in the disk follows Keplerian orbits
one finds that the energy emitted per area $Q$ at radius $r$
is given by (\cite{sha73}):
\begin{equation}
Q = {3 \dot M_{accr} G M_{wd} \over 8 \pi r^3}
    \left[1 - \left({r_o \over r} \right)^{1/2} \right] \;.
\end{equation}
The luminosity of the disk $L_{disk}$ may be obtained by integrating the
energy emission per area $Q$ over the surface of the disk.

Assuming that the disk is emitting locally as a blackbody,
and considering equation~(1),
we have that the radial temperature distribution of the disk will be
\begin{equation}
T = \left\{ {3 \dot M_{accr} G M_{wd} \over 8 \pi r^3 \sigma_s}
    \left[1 - \left({r_o \over r} \right)^{1/2} \right] \right\}^{1/4}
      \; .
\end{equation}
where $\sigma_s$ is the Stefan-Boltzmann constant. 

Equation~(1) is implemented in our model,
taking into account foreshortening and assuming local blackbody
radiation,
in the calculation of radiation flux throughout the wind,
and consequently in the calculation of the continuum and line radiation
pressure force.
Equation~(2) is implemented in our model as a boundary condition for
the wind near the disk surface.
In our model we assume that the temperature at the base of the wind
is equal to the disk surface temperature for each radius.

\section{Line Radiation Pressure Force}

A crucial ingredient of the disk wind models is the description of the
radiation pressure force per mass due to line scattering.
In this section we derive the expression adopted for the
accretion disk wind model.
We obtain a more complex expression than in the isothermal models
we presented in  Paper~I.
The expression obtained here is based upon the radiation emitted by an
accretion disk and includes not only the radiation pressure along the
direction perpendicular to the disk,
but also the radiation pressure parallel to the disk.

Introducing the approximation that the velocity gradient is primarily
in the direction perpendicular to the surface of the disk and
using the opacity distribution developed by Castor, Abbott,
\& Klein~(1975),
we obtain the following expression for the force per mass due to lines
(Paper~I):
\begin{equation}
\vec{f}_{rad} = \oint \left[ \;
         {\sigma \int_0^\infty I(\hat{n}) d\nu \over c } \;
 k \left( { 1 \over \rho \sigma v_{th}}
           \left|\hat{n} \cdot \hat{z} \; {dv_z \over dz}
           \right| \; \right)^\alpha \;
   \right] \hat{n} d\Omega \; ,
\end{equation}
where $\sigma$ is the Thomson cross section per unit mass,
$I$ the specific intensity,
$c$ the speed of light,
$k$ and $\alpha$ are parameters dependent on temperature and the
relative abundance of different elements,
$\rho$ is the mass density,
$v_{th}$ is the thermal sound speed,
and $v_z$ is the component of the wind velocity in the $z$ direction.

We wish to note here that although equation~(3) takes into account the
line radiation pressure in the ``$r$'' direction,
we have included only the velocity gradient in the ``$z$'' direction.
As we will show in \S~4,
the accretion disk winds in CVs tend to flow in the direction
perpendicular to the disk,
particularly in the region near the surface of the disk,
thus justifying the assumption of a
velocity gradient primarily in the ``$z$'' direction.
Although on one hand, incluying the velocity gradients en the ``$r$''
direction may possibly produce interesting results, on the other we do not
believe that it will significantly affect the overall results.
Also, as we show below, the assumption of a velocity gradient primarily
in the ``$z$'' direction allows us to obtain an
expression for the line radiation pressure that requires the
calculation of a spatial integration only once,
rather than having to evaluate the spatial integrals at each time step,
thus making the computations manageable.

Abbott~(1982) presented elaborate calculations for the $k$ and
$\alpha$ parameters considering temperatures varying from $6,000$
to $50,000 \; {\rm K}$. He found that the $k$ parameter to be generally
an order of magnitude greater than the earlier values presented by
Castor, Abbott, \& Klein (1975) in which the line radiation parameters
calculations were based on atomic line data of the \ion{C}{3}.
Thus for the models presented in this paper we have used the values
$k=1/3$ and $\alpha=0.7$, rather than the earlier values used by
Castor, Abbott, \& Klein (1975) of $k=1/30$ and $\alpha=0.7$ as we
did in our isothermal models presented in Paper~I.
Thus we feel confident that our formulation is approximately correct
for disk temperatures throughout this range.
We wish to note here that Abbott~(1982) found values for $\alpha$ which
varied roughly between $0.5$ and $0.7$, and that as we have observed
through our isothermal models (Paper~1)
values of $\alpha$ near $0.5$ will reduce significantly both the wind mass
loss rate and the wind velocity. In our next paper we will implement
local ionization equilibrium and calculate the $k$ and $\alpha$ parameters
at each point of the wind rather than assuming fiducial values for these
parameters throughout the wind as we have done up to now.
Note that the parameter $\alpha$ used here and throughout this work is
not the same as the parameter often used to describe accretion disk
viscosity (e.g., \cite{sha73}).

From equation~(3) we evaluate the angular dependence exactly,
rather than introduce approximations for the angular dependence
as we did for our isothermal models presented in Paper~1.
Therefore:
\begin{equation}
\vec{f}_{rad} = {\sigma \over c} \;
                k \left( { 1 \over \rho \sigma v_{th}}
                \left| {dv_z \over dz} \right| \right)^\alpha 
                \oint
                \left[ \; \left|\hat{n} \cdot \hat{z} \right|^\alpha 
                {\int_0^\infty I(\hat{n}) d\nu} \;
                \right] \hat{n} d\Omega \; .
\end{equation}
Defining $\vec S(r,z)$:
\begin{equation}
\vec S(r,z) = \oint
              \left[ \; \left|\hat{n} \cdot \hat{z} \right|^\alpha 
              {\int_0^\infty I(\hat{n}) d\nu} \;
              \right] \hat{n} d\Omega \; ,
\end{equation}
Taking the source of radiation to be the accretion disk,
we have
\begin{equation}
S_z(r,z) = \int_{r_o}^\infty \int_0^{2\pi}
           {Q(r') \over \pi} \; {z^{\alpha+2} \over
            [(r^2+r'^2+z^2-2rr'\cos\phi)^{1/2}]^{\alpha+4} }
           \; r' d\phi dr'
\end{equation}
and
\begin{equation}
S_r(r,z) = \int_{r_o}^\infty \int_0^{2\pi}
           {Q(r') \over \pi} \; {z^{\alpha+1}(r-r'\cos\phi) \over
            [(r^2+r'^2+z^2-2rr'\cos\phi)^{1/2}]^{\alpha+4} }
           \; r' d\phi dr' \;
\end{equation}
where $S_z$ and $S_r$ are the corresponding components of the vector
$\vec S$ defined in equation~(5)
(the $S_\phi$ component is zero due to the axial symmetry of the
accretion disk)
and $Q(r')$ is the radiation emission per area of the disk calculated
through equation~(1).

Thus we have 
\begin{equation}
\vec{f}_{rad} = {\sigma \over c} \;
                k \left( { 1 \over \rho \sigma v_{th}}
                \left| {dv_z \over dz} \right| \right)^\alpha 
                [ S_z(r,z) \hat z + S_r(r,z) \hat r]
\end{equation}
where $S_z$ and $S_r$ are given by equations~(6) and (7).

We also account for the fact that,
as can be seen from the Appendix of Paper~1,
if the velocity gradient increases to sufficiently high values,
or the density of the wind decreases to sufficiently low values,
the contribution for the radiation pressure due to lines arrives at a
maximum value.

Abbott (1982) found that the maximum value of the line radiation
pressure was obtained when
$(1/\rho \sigma v_{th}) dv_z/dz \lesssim 10^7$.
In this work we have assumed that the maximum possible value of the
line radiation pressure is obtained when
$(1/\rho \sigma v_{th}) dv_z/dz = 10^8$.
We note that we may be overestimating the maximum value of 
$(1/\rho \sigma v_{th}) dv_z/dz$, and thus the maximum value
of the force multiplier.
We do not expect this to cause a significant difference
in the overall dynamics,
particularly in the high density regions where most of the
spectral lines are forming.
This will be discussed again in \S~4.3~.
In the future we shall include local ionization equilibrium into
our models self consistently and calculate the line radiation force
multiplier at each point in the wind in a manner equivalent
to Abbott's calculation for OB stars (\cite{abb82}).

Therefore the expression for the total line radiation force per mass
for disk winds which we adopt is
\begin{equation}
\vec{f}_{rad} = {\sigma \over c} \;
                k \left( \max \left[{ 1 \over \rho \sigma v_{th}}
                \left| {dv_z \over dz} \right|\; ,\; 10^8 \right] \;
                \right)^\alpha 
                [ S_z(r,z) \hat z + S_r(r,z) \hat r] \; .
\end{equation}

For the continuum radiation pressure the total radiation flux is
calculated throughout the disk wind
\begin{equation}
\vec F = \oint \left[ \; {\int_0^\infty I(\hat{n}) d\nu} \;
                \right] \hat{n} d\Omega \; .
\end{equation}
Taking the source of radiation to be the accretion disk, we have
\begin{equation}
F_z(r,z) = \int_{r_o}^\infty \int_0^{2\pi}
           {Q(r') \over \pi} \; {z^2 \over
            [(r^2+r'^2+z^2-2rr'\cos\phi)^{1/2}]^4 }
           \; r' d\phi dr'
\end{equation}
and
\begin{equation}
F_r(r,z) = \int_{r_o}^\infty \int_0^{2\pi}
           {Q(r') \over \pi} \; {z (r-r'\cos\phi) \over
            [(r^2+r'^2+z^2-2rr'\cos\phi)^{1/2}]^4 }
           \; r' d\phi dr' \;
\end{equation}
where $F_z$ and $F_r$ are the corresponding components of the radiation
flux vector $\vec F$ originating from the disk (the $F_\phi$ component
is zero due to the axial symmetry of the accretion disk) and $Q(r')$ is
the radiation emission per area of the disk calculated through
equation~(1).

The line radiation pressure depends strongly on the radiation flux
(equations~[5] and [8]),
and therefore the radiation flux plays an important role in our models.
In Figure~3 we present the radiation flux perpendicular to the disk as
a function of height for different radii.
A peculiarity of the disk winds is that gravity increases along the
streamlines.
In our accretion disk wind models for large radii
(see Figure~3)
we find an increase in radiation flux with height along the streamlines
also.
This is necessary for wind driving.
For small radii
($ r \lesssim 5 R_{wd}$)
(see Figure~3)
we observe little or no increase in radiation flux with height,
but we do observe a significant increase in the radiation flux with
decreasing radius
($\approx$ a factor of $10^6$ between the innermost [$2R_{wd}$] and
outermost [$150R_{wd}$] radii found in the spatial grids of our
computational models).
For the inner disk radii ($r = 2 R_{wd}$) the radiation flux is
locally super-Eddington,
and the gravity increase is overcome by the high radiation flux.

\section{Hydrodynamic Model}

\subsection{Equations}

Our hydrodynamic model uses the Piece-Wise Parabolic Method (PPM)
numerical scheme (\cite{col84}).
We use the 2.5-dimensional hydrodynamic equations of the wind in
cylindrical coordinates.
That is,
the three-dimensional hydrodynamic equations are reduced to
two-dimensional equations by assuming that the derivative of any
physical variable with respect to $\phi$ is zero.
The first spatial dimension is the height above the disk $z$.
The second spatial dimension is the distance between the center of the
disk and the projection on the disk $r$.

The equations are:
the equation of state,
\begin{equation}
P = (\gamma -1 )\rho e \; ,
\end{equation}                  
the mass conservation equation,
\begin{equation}
{\partial\rho \over \partial t}
  + {1 \over r}{\partial(r \rho v_r) \over \partial r}
  + {\partial(\rho v_z) \over \partial z} = 0 \; ,
\end{equation}
the momentum conservation equations,
\begin{eqnarray}
\rho{\partial v_r \over \partial t}
  + \rho v_r{\partial v_r \over \partial r}
  - \rho {{v_\phi}^2 \over r}
  + \rho v_z{\partial v_r \over \partial z}
& = &
  -   \rho{GM_{wd} \over (r^2 + z^2)}{r \over (r^2+ z^2)^{1/2}} 
  - {\partial P \over \partial r} + \rho {\sigma F_r(r,z) \over c}
    \nonumber \\
&   & \\
&&+   \rho {\sigma S_r(r,z) \over c} \; 
      \times
         k \left( \max \left[{ 1 \over \rho \sigma v_{th}}
         \left| {dv_z \over dz} \right|\; ,\; 10^8 \right] \;
         \right)^\alpha  \; , \nonumber 
\end{eqnarray}
\begin{equation}
\rho {\partial v_\phi \over \partial t}
  + \rho v_r{\partial v_\phi \over \partial r}
  + \rho {v_\phi v_r \over r}
  + \rho v_z{\partial v_\phi \over \partial z}
  = 0 \; ,
\end{equation}
\begin{eqnarray}
\rho{\partial v_z \over \partial t}
  + \rho v_r{\partial v_z \over \partial r}
  + \rho v_z{\partial v_z \over \partial z}
& = &
 - \rho{GM_{wd} \over (r^2+z^2)^2}{z \over (r^2+z^2)^{1/2}} 
 - {\partial P \over \partial z} + \rho {\sigma F_z(r,z) \over c} 
   \nonumber \\
&   & \\
&& + \rho {\sigma S_z(r,z) \over c} \; 
  \times k \left( \max \left[{ 1 \over \rho \sigma v_{th}}
         \left| {dv_z \over dz} \right|\; ,\; 10^8 \right] \;
         \right)^\alpha  \; , \nonumber
\end{eqnarray}
and the energy conservation equation,
\begin{eqnarray}
{\partial \rho E \over \partial t}
+ {1 \over r}{\partial r \rho E v_r \over \partial r}
+ {\partial \rho E v_z \over \partial z}
& = &  - {1 \over r}{\partial r P v_r \over \partial r}
  -{\partial P v_z \over \partial z} \nonumber \\
&   &  \nonumber \\
&& - \rho{GM_{wd} \over (r^2 + z^2)}{r \over (r^2+ z^2)^{1/2}} v_r 
 + \rho {\sigma F_r(r,z) \over c} v_r \nonumber \\
&   &  \\
&& + \rho {\sigma S_r(r,z) \over c} \; 
  \times k \left( \max \left[{ 1 \over \rho \sigma v_{th}}
         \left| {dv_z \over dz} \right|\; ,\; 10^8 \right] \;
         \right)^\alpha v_r \nonumber \\
&   &  \nonumber \\
&& - \rho{GM_{wd} \over (r^2+z^2)^2}{z \over (r^2+z^2)^{1/2}} v_z 
 + \rho {\sigma F_z(r,z) \over c} v_z  \nonumber \\
&   &  \nonumber \\
&&  + \rho {\sigma S_z(r,z) \over c} \; 
  \times k \left( \max \left[{ 1 \over \rho \sigma v_{th}}
         \left| {dv_z \over dz} \right|\; ,\; 10^8 \right] \;
         \right)^\alpha v_z \; , \nonumber
\end{eqnarray}
where $P$ is the pressure,
$\gamma$ is the ratio of specific heats,
$\rho$ is the density,
$e$ is the internal energy per mass,
$v_r$,
$v_\phi$,
and $v_z$ are the corresponding velocity components in cylindrical
coordinates,
$G$ is the gravitational constant,
$M_{wd}$ is the mass of the white dwarf,
$\sigma$ is the Thomson cross section per unit mass,
$F_r$ and $F_z$ are the corresponding radiation flux components
(see equations~[11] and [12]),
$c$ is the speed of light,
$v_{th}$ is the ion thermal velocity,
and $k$ and $\alpha$ are defined through equation~(3),
$S_r$ and $S_z$ are defined through equations~(7) and (6)
respectively,
and $E = v_r^2/2 + v_\phi^2/2 + v_z^2/2 + e$ is the total energy per
mass.

Equations~(14)-(18) are solved with the PPM scheme for
cylindrical coordinates.
We wish to note here that our implementation of the PPM scheme in
two dimensions in cylindrical coordinates
(``z'' and ``r'')
does not include the third term on the left hand side of equation~(16),
and that we have included and implemented this term in our models
explicitly.
We note also that the centrifugal and coriolis forces
appear naturally under cylindrical coordinates as the third term of the
left hand side in equation~(15) and the third term of the left hand
side in equation~(16).

In this treatment we have implemented the energy equation self
consistently.
In particular,
the adiabatic heating and cooling effects are implemented through the
first and second term on the right hand side of equation~(18).

The emission of the boundary layer is one of the major unsolved problems
in the study of cataclysmic variables,
and it is very closely related to the emission from the inner regions
of the accretion disk.
Owing to the observational uncertainties discussed in the Introduction,
we have chosen not to include the boundary layer radiation in our
current models.
In the model presented here we solve the hydrodynamic equations over a
range of radii from $0.02 R_{\sun}$
($2R_{wd}$ in our models)
to $1.5 R_{\sun}$. We wish to note at this point that we have
run additional models over a range of radii from $0.01R_{\sun}$
($1R_{wd}$ in our models)
to $0.3 R_{\sun}$ finding similar results,
comparisons are presented below.
From equation~(1) it is found that the energy emission per
area $Q$ will increase slightly as radius decreases from
$2R_{wd}$ down to $\approx 1.36 R_{wd}$  and will decrease
strongly as radius decreases from $\approx 1.36 R_{wd}$
down to the white dwarf radii ($R_{wd}$).
The function $Q(r)$ (equation~[1]) was originally derived by
Shakura and Sunyaev (1973) for binary systems with an accretion disk
about a black hole.
Popham and Narayan (1995) developed detailed models of boundary layers
and accretion disks in cataclysmic variables and found that their
models supported the same function $Q(r)$ for the accretion disk
in CVs.
In Paper~I we showed that the wind density tends to decrease
with decreasing radiation flux.
Thus the the mass loss rate for radii less that $2 R_{wd}$ will not
alter significantly the total mass loss rate of the wind.

\subsection{Boundary Conditions}

As we discovered in Paper~I,
care must be taken with the boundary conditions to ensure not only a
numerically stable solution but also a physically stationary one.
The PPM scheme (\cite{col84}), in order to obtain the physical
parameters (velocity, density, pressure) of the wind at a
given point of the computational grid $x_i$ for a given time step
$t_j+\Delta t_j$,
requires that at time $t_j$ the values of the physical parameter
be known at the given spatial grid point ($x_i$),
at the following two grid points ($x_{i+1}$ and $x_{i+2}$),
and at the previous two grid points ($x_{i-1}$ and $x_{i-2}$).

Therefore, the boundary condition in PPM must be implemented by
not only specifying the values of the physical parameters
(or leaving them free depending on the specific boundary condition)
at each grid point in the boundary of the spatial computational grid,
but also by specifying the physical parameters at two grid points
beyond each boundary grid point.
The latter set of two spatial grid points are named ``ghost zones''
since they are points off the actual computational grid.

The lower boundary condition of our model is implemented in the
following manner.  The lower boundary of our computational grid is
set at $z_0=0.023R_{\sun}$,
which corresponds to the photospheric height at $r=R_{\sun}$,
and which we adopt as a fiducial value over the range of radii
of our computational grid
(\cite{sha73}, Paper~I).
The velocity along the $z$ direction is fixed at
$v_z=10 \; {\rm km \; s}^{-1}$,
which is a fiducial value of the sound speed at the photospheric
height over the range of radii implemented in our model.
We note that in the inner regions of the disk ($r < 0.25R_{\sun}$),
where the temperature will be greater that $6,000~{\rm K}$
(equation~[2]),
our boundary conditions will be subsonic.
Since the radial forces balance on the disk surface,
the initial wind velocity must be normal to the surface of the disk,
therefore the velocity along the $r$ direction is fixed at a value of
$v_r=0$. The velocity along the $\phi$ direction is fixed at Keplerian
speeds.
The density is left free.
The pressure at the lower boundary is calculated at each time step
according the current the value of density and temperature.
The temperature is fixed according to equation~(2).

The photospheric height at the innermost radius of our computational
grid
($r=0.02 R_{\sun}$)
is actually two orders of magnitude less than the photospheric height
at $r=R_{\sun}$
(\cite{sha73}).
We have run our models varying the fiducial photospheric height
(lower computational grid boundary) from a value of
$z_0=0.023R_{\sun}$
down to a value of
$z_0=0.00023R_{\sun}$
and found similar results for lower computational boundaries
within that range (see Figure~11).

The ghost zones for the lower boundary are set in the following manner.
The velocity along the $z$
direction are fixed according to the following equation:
\begin{equation}
v_z = v_{0z}*e^{-{1 \over 2}}
     *e^{{1 \over2}(z / z_0)^2}
 \; , \; (z < z_0) \; ,
\end{equation}
where $v_{0z}$ is the velocity at the lower boundary
($10 \; {\rm km \; s}^{-1}$).
Equation~19 was obtained by an approximate integration of the
hydrodynamic equations,
in the subsonic region,
of the one-dimensional models we presented in Paper~1.
Zero gradients are implemented for the temperature and for the
velocities in the $r$ and $\phi$ directions.
The density in the ghost zones is calculated such that
$\rho v_z = {\rm constant}$, to ensure mass conservation.
The pressure is calculated at each time step according to the current
value of density and temperature.

For the left, upper, and right boundaries the physical parameters are
left free at each grid point in the computational boundary.
Zero gradients are implemented at the ghost zones for the physical
parameters.

\subsection{Intermediate Models}

The models presented in this paper include several important
computational improvements and changed physical assumptions
when compared with Paper~1.
In order to understand the effects of these separately we present
results of intermediate models.

As we already indicated in \S~3,
in the accretion disk wind model presented here,
for the force multiplier parameters,
we have taken values presented by Abbott~(1982)
$k=1/3$ and $\alpha=0.7$.
In our isothermal models presented in Paper~1, 
for the force multiplier parameters we took the earlier
values used by Castor, Abbott, and Klein~(1975) $k=1/30$ and
$\alpha=0.7$.
In Figure~4 we show results from our isothermal models presented in
Paper~1,
in Figure~5 we show results from a similar model which differs only in
the implementation of the higher and more recent value for the
$k$ parameter.
We note that in both models the velocity fields are essentially the
same.

The mass density distribution is also quite similar except for a factor
of $\approx 27$ in the density values of the newer models.
From equation~3 we note that the increase in the value of
$k$ is equivalent
to an increase in luminosity
(with respect to line-radiation pressure).
In turn from the one-dimensional models presented in Paper~1
we showed that,
in analogy with early type stars,
an increase in luminosity essentially leaves unaltered the wind
velocities and increases the overall density in proportion to
$L^{1/\alpha}$.
Thus our models in Paper~1 indicate that an increase in a factor
of $10$ in the $k$ parameter should produce an increase in
overall density by a factor of $10^{1/0.7} \approx 27$ which is
consistent with the results presented in Figure~5~.

As seen in equation~(9),
in the accretion disk wind model presented here we have also
taken into account the fact that if the velocity gradient increases
to sufficiently high values,
or the density decreases to sufficiently low values,
the contribution for the radiation pressure due to lines arrives
at a maximum value.
In Figure~6 we show the results from a model in which we implement
this condition.
The models represented in Figures~5 and 6 differ only in that
former model does not implement a maximum value for the line
radiation pressure while the latter does.

We wish to note here two of the effects we find in implementing
a maximum value for the force multiplier.
First,
in the region high above the inner disk area
(upper left region in Figures~5 and 6)
a decrease in velocity is obtained.
We find that the ``z'' velocities components in the fore mentioned
region in Figure~5
($\gtrsim 10,000 {\rm \; km \; s}^{-1}$)
are higher than the respective velocities in Figure~6
($\sim 1,000 {\rm \; km \; s}^{-1}$).
This difference is due to the fact that the region high above
the inner disk area is a low density region since centrifugal
forces have driven most of the mass towards larger ``$r$''
(towards the right in Figures~5 and 6)
at the corresponding heights
($z \gtrsim 2.5 R_{disk}$).
In turn the decrease in density in this region produces an
increase in the line radiation pressure (see equations~[3] and [9])
which is allowed to increase indefinitely in the model of Figure~5,
but may only increase up to a maximum value in the model
of Figure~6.
Second,
we also find that the implementation of a maximum value for the force
multiplier also produces an increase in density in the region near the
base of the wind by a factor of $\sim 4$~.

In our accretion disk wind model we also introduced an adiabatic
wind rather than an isothermal wind as we assumed in Paper~1.
In Figure~7 we show results from a model which implements
an adiabatic wind.
The models represented in Figures~6 and 7 differ only in that
former model assumes an isothermal wind while the latter assumes
an adiabatic wind.
We do not find significant changes in the density distributions or
the wind velocity field when we implement an adiabatic wind
rather than an isothermal wind
(see Figures~6 and 7). 

Another element that we have introduced in our
accretion disk wind model is the radial structure of
radiation emission 
(equation~[1])
and temperature of disk surface
(equation~[2]),
rather than assume an isothermal disk as we did in the
models presented in Paper~1.
In Figure~8 we show results from our accretion disk wind model.
The models represented in Figure~7 and Figure~8 differ in that
former assumes an isothermal disk
(and therefore a uniform temperature and radiation emission
distribution over the disk surface)
and neglects the radiation pressure along the ``$r$'' direction,
while the latter assumes a radial temperature and radiation
emission distribution as described in \S~2 and includes
the radiation pressure along the ``$r$'' direction.

Comparing the results shown in Figures~7 and 8,
we note that the enhanced density region above the disk
extends from the mid-region of the disk in Figure~7,
while in Figure~8 the enhanced density region extends from the
inner region of the disk.
This difference is due to the assumption of an isothermal disk
in the Figure~7 model which produces a uniform density distribution at
the base of the wind,
while in Figure~8 the higher radiation emission from the inner
disk region (equation~[1]) produces in turn a higher density
at the inner region  than in the outer region at the base of
the wind.

The radial radiation emission structure of the accretion disk wind
model causes the density in the inner region of the disk at the base
of the wind to be higher than the corresponding
density at the same region in the isothermal disk model
(see Figures~7 and 8).
This in turn causes an increase in the overall wind mass loss rate
by a factor of $\sim 4$ in the accretion disk wind model
(Figure~8)
with respect to the isothermal disk wind model
(Figure~7).

The results shown in Figure~8 allow us to estimate the validity of
our neglect of radiative cooling.
We present here rough estimates of the radiative cooling times and
compare them with dynamical time scales.
Blondin (1994) obtained an analytical expression for radiative cooling
in high-mass X-ray systems.
He found for wind temperatures on the order of $T \sim 10^4 {\rm K}$
a radiative cooling coefficient of
$\Lambda \approx 10^{-24} {\rm erg \; cm}^3 {\rm s}^{-1}$.
Rough estimates for the cooling times may be obtained through
the following expression $t_{cool} \sim k T / n \Lambda$,
where $k$ is boltzmann's constant, $T$ is the temperature,
and $n$ is the particle density.
From the results of Figure~8 (discussed below) we obtain the following
rough estimates for the radiative cooling times:
lower inner region ($r=0.3R_{\sun}$, $z=0.2R_{\sun}$)
$t_{cool} \sim 640 {\rm \; s}$;
lower outer region ($r=0.8R_{\sun}$, $z=0.2R_{\sun}$)
$t_{cool} \sim 2.6 \times 10^{5} {\rm \; s}$;
higher inner region ($r=0.3R_{\sun}$, $z=2.5R_{\sun}$)
$t_{cool} \sim 4.6 \times 10^{7} {\rm \; s}$;
higher outer region ($r=0.8R_{\sun}$, $z=2.5R_{\sun}$)
$t_{cool} \sim 3,8 \times 10^{3} {\rm \; s}$.
A rough estimate of the dynamical time maybe obtained by
$t_{dynamic} \sim 4*R_{\sun}/4,000 {\rm km \; s}^{-1}$,
$t_{dynamic} \sim 700 {\rm \; s}$.
Viewing Figure~8 we note that radiative cooling time in the
high density areas tend to be comparable to or larger than
the dynamical time.
This suggests that in some parts of this region,
radiative cooling may be important and the EOS could be described as
isothermal.
In the lower density areas the radiative cooling time scales
tend to be much larger than the dynamic time scale,
thus indicating that for this region the EOS would be best described as
adiabatic.
These estimates are uncertain by factors of an order of magnitude
or more,
owing to the fact that they assume optically thin line cooling and are
not tailored to the conditions specific to CV winds.
However,
they provide motivation for our future work,
in which we shall include radiative heating and cooling in our models.

\subsection{Results}

The spatial computational grid in the $z$ direction has $101$ points,
ranging from the photospheric height to $61 R_{\sun}$,
distributed such that $\arctan (z_i/R_{\sun})$ is uniform.
The number of grid points for the spatial computational grid in the
$r$ direction was set at different values which varied from $11$ to
$1001$ distributed uniformly.
We obtain similar results for each of the grid resolutions implemented
within the above range.

A vector field graph of wind velocity superimposed with a contour
graph of wind density is presented in Figure~8 for the low resolution
model.
A vector field graph of wind velocity superimposed with a contour
graph of wind temperature is presented in Figure~9~.
The physical parameters used are: $M_{wd} = 0.6M_{\sun}$,
$R_{wd} = 0.01 R_{\sun}$,
and $L_{disk} = L_{\sun}$.
We obtain the following values for the mass-loss rate and terminal
velocity for our two-dimensional accretion disk wind model:
$\dot M_{wind}=8 \times 10^{-12} M_{\sun} {\rm yr}^{-1}$
(obtained by integrating at the photospheric height)
and $v_\infty \sim 3000 \; {\rm km} \; {\rm s}^{-1}$.

We note that the wind mass loss rate we find for our accretion disk
wind model is approximately two orders of magnitude larger
than the value we found in Paper~I for our isothermal models.
This is due to three reasons:
first,
the radiation field above the inner region of disk in the accretion
disk wind models presented in this work is stronger than in the models
presented in Paper~I in which we assumed an isothermal disk.
Second,
we have used a more recent and larger value for the $k$ force
multiplier parameter in our accretion disk wind model.
For the line radiation pressure,
the effect of increasing $k$ is equivalent to increasing luminosity
(see equations~[6], [7], and [9]).
As we had shown in Paper~I,
through our one-dimensional isothermal models,
an increase in disk luminosity produces a corresponding increase in
wind mass-loss rate.
Third,
in our accretion disk wind model we implement the fact that,
if the velocity gradient increases to sufficiently high values
or the density of the wind decreases to sufficiently low values,
the contribution for the radiation pressure due to lines arrives at a
maximum value (see equation~[9]).
This causes a decrease in the line radiation pressure in the regions of
the wind where this phenomena occurs.
Of these three factors, the second, the use of a larger and more recent
value for $k$, is the most important, accounting for slightly more than
a order magnitude increase in the wind mass loss rate.

In Figure~10 we show results obtained from the low resolution model
($11$ computational grid points in the $r$ direction)
and from the high resolution model
($1001$ computational grid points in the $r$ direction),
and as stated above we find similar results
(the density contour values on both Figures~10{\it a} and 10{\it b}
are the same).
We note structure in the density contours in the  $1001$ computational
grid which is not apparent in the density contours in the
$11$ computational grid.
Figures~10{\it a} and 10{\it b} are shown at the same physical times,
in which steady flows have been obtained.
The motivation for the high resolution models was to ensure that the
overall results of our low resolution models were not affected by the
selection of vertical computational grid.
Comparing Figures~10{\it a} and 10{\it b} we note structure in the
density contours in the  $1001$ computational grid which is not
apparent in the density contours in the $11$ computational grid.
We have not yet fully explored the density spikes in the high resolution
model.
However,
the average properties of the flow such as total mass loss rate and
gross structures,
are equivalent in the two models.

In Figure~11 we show results obtained from a model with
a height of $z_0=0.00023R_{\sun}$ at the lower computational
grid boundary
(Figure~11{\it b})
and compare it with the results from a model with a
height of $z_0=0.023R_{\sun}$ at the lower computational
grid boundary 
(Figure~11{\it a})
($z_0=0.023R_{\sun}$ is the lower computational boundary used
in the models presented elsewhere in this paper).
The density contour values on both Figures~11{\it a}
and 11{\it b} are the same.
As we indicated in \S~4.2,
these two heights correspond to the photospheric height in the innermost
radius of our computational grid ($r=0.02R_{\sun}$) and to the
photospheric height in the radius of $r=R{\sun}$ respectively.

In the models with lower photospheric heights we did not increase
the number of grid points in the photospheric region.
The motivation for this is that in the region between the original
photospheric height and the smaller values,
the radiation field is not changing considerably.
This, added to the consideration that this region
is appreciably smaller than the total spatial computational grid,
would indicate that a finer vertical grid should produce no
significant difference, while considerably increasing the CPU time
required for the actual runs.

In Figure~12 we present a vector field graph of wind velocity
superimposed with a contour graph of density for the
accretion disk wind model with the same physical parameters,
but implementing a reflecting wall boundary condition for the left
side of the computational grid.
That is,
the values for the physical parameters
(velocity, density, and pressure)
are left free at the grid points in the left computational
boundary, and are therefore recalculated at each time step.
``Mirrored'' values are implemented at the ghost zones for the
physical parameters for the reflecting wall boundary conditions.
With the reflecting wall boundary condition we obtain quite similar
results
(see Figures~8 and 12).

We wish to note that in Figure~12 the ``r'' components of the velocity
in the upper region at the inner left boundary do not indicate
that there is a mass flow coming from the inner left boundary,
but rather that the gas pressure is overcome by centrifugal forces
in this region.
We also wish to note that in the reflecting wall boundary condition
the grid points on the left computational boundary are not
at the ``wall'',
but rather near the ``wall''.
The ``wall'' would be located in between the boundary grid points
and the ``ghost zone'' grid points (see \S~4.2).

In Figure~13 we show results obtained from the low resolution model
(Figure~13{\it a}) and two additional high resolution models
(Figures~13{\it b} and 13{\it c}).
As stated above the low resolution model contains $11$ computational
grid points in the $r$ direction ranging from $0.02 R_{\sun}$ ($2.0 R_{wd}$)
to $1.5 R_{\sun}$, and $101$ grid points in the $z$ direction ranging from
$0.023 R_{\sun}$ to $61.0  R_{\sun}$. In this figure we show the
results of the low resolution model up to $0.3 R_{\sun}$ in both directions
in order to present them at the same scale as Figures~13{\it b}
and 13{\it c}.
The high resolution models represented in Figure~13{\it b} and 13{\it c}
have $1001$ grid points in the $r$ direction ranging up to $0.3 R_{\sun}$,
and $501$ grid points in the $z$ direction ranging from $0.023 R_{sun}$
to $0.3 R_{\sun}$. Figures~13{\it b} and 13{\it c} differ in that in the
former the grid points in the $r$ direction range down to $0.02 R_{\sun}$
($2.0 R_{wd}$), while in the latter the grid points in the $r$ direction
range down to $0.01 R_{\sun}$ ($1.0 R_{\sun}$).
As we stated above we find similar results between the low and the high
resolution models.
The velocity fields are essentially the same and the density in the
corresponding region are of the same magnitude. Although we do find more
structure in the density contours of the high resolution models,
which are not apparent in the low resolution models. In the high resolution
models we also find that near the disk surface towards larger radii
lower densities are obtained.
The motivation for the additional high resolution models was to ensure
that the overall results of our models would not be affected by a more
detailed modeling of the wind intiation region through the increase of
the physical resolution on one hand and by including in the models
the region between $1.0R_{wd} \leq r \leq 2.0R_{wd}$ on the
other. We wish to note here that in the Figures~13{\it a}, 13{\it b},
and 13{\it c} the velocity vectors presented with the lowest $z$ are the
lowest $z$ in which the velocity vectors are plotted, the actual
computational grid starts at a lower $z$ as described above in this
paragraph.

An important feature of the two-dimensional accretion disk wind
model results,
the role of rotational forces, is apparent in Figures~8 and 9~.
The centrifugal forces cause the streamlines from the inner region of
the disk to bend and to collide with streamlines of material
originating at larger radii.
This collision generates a large region of enhanced density which runs
diagonally in the $z$-$r$ plane
(this effect was also observed in Paper~I for our two-dimensional
isothermal models).
An important consequence of the enhanced density region is the
corresponding increase in the line optical depth.

Icke~(1981) suggested that biconical winds were a general property
of accretion disk winds independent of the wind driving mechanism,
and that this was caused by the geometry of the disk wind and by
the rotation of the disk.
In this work we have confirmed this suggestion.
We find that the principal driving mechanism for accretion disk winds
in CVs is the line radiation pressure and that for our
two-dimensional models we find that centrifugal forces,
which are caused by the rotation of the disk,
produce an enhanced density region which is biconical in nature
(Figure~8).

\section{\ion{C}{4} 1550 \AA \ Line Profiles}

In order to further study line-driven accretion disk winds we calculate
\ion{C}{4}~1550~\AA \ line profiles from our accretion disk wind
model.
We then study the line profiles calculated at different angles and
compare them with observations.

The line profiles presented here (Figure~14) were calculated assuming
single scattering and a relative abundance of carbon with respect to
hydrogen of $n_{\rm CIV}/n_{\rm H} = 10^{-3}$ throughout the wind.
The radiation was assumed to have originated on the surface of the
disk,
according to equation~(1),
in consistency with our model.
The angles for the line of sight with respect to the disk rotation
axis were discretized into eighteen values:
from $0^\circ$ to $85^\circ$ in $5^\circ$ increments.
We wish to note here that although the three-dimensional
hydrodynamic equations were reduced to and implemented in
two dimensions through azimuthal symmetry (2.5-dimensional),
the line profiles were calculated in three dimensions.

For a given point source on the disk surface and a given direction,
the geometric line starting from the given point and having the given
direction,
was divided into several segments as it intersected the spatial
computational grid.
A three-dimensional computational grid was formed by rotating the
spatial grid used in the hydrodynamic calculations along the $\phi$
direction.
Continuous values of $\phi$ were used to calculate the intersections.
Over each of these segments the optical depth for each frequency bin,
corresponding to a given velocity bin,
was obtained by taking into account the ion thermal speed $v_{th}$ and
applying the expression
$\tau = \int \rho \kappa_L ds$ along a Sobolev length
(a Sobolev length is defined as the distance along which the wind
velocity varies by a value equal to the sound speed).
That is, for a given point source on the disk surface and a
given direction, the following equations are implemented:
\begin{equation}
\tau_\nu = \int\limits_{v-v_{th}/2}^{v+v_{th}/2} \rho \kappa_L ds
\; ,
\end{equation}
\begin{equation}
\nu = {\nu_0 \over   1 - v/c } \; ,
\end{equation}
\begin{equation}
I_\nu ({\rm non-scattered}) = S_\nu \, e^{-\tau_\nu} \; ,
\end{equation}
\begin{equation}
I_\nu ({\rm scattered}) = S_\nu \, (1-e^{-\tau_\nu}) \; ,
\end{equation}
where $\tau_\nu$ is the optical depth for frequency $\nu$,
$v$ is the component of the wind velocity along the given direction,
$v_{th}$ is the thermal speed,
$\rho$ is the wind density,
$\kappa_L$ is the line opacity for \ion{C}{4} 1550~\AA,
$\nu_0$ is the resonance frequency for \ion{C}{4} 1550~\AA,
$c$ is the speed of light,
$I_\nu$ is the intensity for frequency $\nu$ for
non-scattered radiation along the given direction,
and $S_\nu$  is the intensity for frequency $\nu$
along the given direction at the disk surface.
We note here that the above calculations were done in the vicinity
of the \ion{C}{4} 1550~\AA \ resonance frequency
in order to calculate the corresponding line profiles.

The total optical depth,
for a given point source on the disk surface,
for a given direction,
for a given velocity bin,
was calculated by summing over the contributions of each segment
described above.
In the line profiles calculations presented here it was
assumed that the radiation absorbed at each segment is
scattered isotropically, and to therefore contribute
uniformly to the radiation in each direction.
For each direction of scattered radiation,
the velocity of the wind was taken into account through
Doppler shifts, 
and then added on to the corresponding velocity bin.
The velocity component along the $\phi$ direction was taken into
account consistently throughout the line profile calculations.

Finally,
after implementing all of the corresponding integrals,
the contribution of the scattered and non-scattered radiation for each
velocity bin was summed for each observation angle and the line profiles
obtained.

The velocity values were discretized in bins of
$100 \; {\rm km} \; {\rm s}^{-1}$.
Although this bin size is greater that the ion thermal speed
($v_{th} \sim 10 {\rm \; km \; s}^{-1}$)
it is much less than the wind velocities and emission line widths that
characterize \ion{C}{4} line profiles in CVs
($\sim 2,000 {\rm \; km \; s}^{-1}$).

For low inclination angles ($0^\circ \leq i \lesssim 55^\circ$)
our model predicts P~Cygni profiles.
The blue-shifted absorption line presents a maximum absorption at
roughly half the terminal velocity,
which is consistent with observations
(\cite{dre87}; \cite{kal88}).
This feature distinguishes the P~Cygni line profiles from CVs
from the P~Cygni line profiles seen from early type stars whose
blue-shifted absorption line presents a maximum absorption at
the terminal velocity (\cite{kal88}).
Furthermore the absorption line corresponds to velocities in the order
of $3000 \; {\rm km} \; {\rm s}^{-1}$,
which also agrees with observations.
For example,
the line profile in Figure~16 shows a strong similarity to the line
profiles observed from RW~Sex (\cite{gre82}, \cite{vit93}, \cite{pri95})
which has an inclination angle of $i \sim 35^\circ$
(\cite{pri95}).

For low inclination angles close to the orbital axis
($0^\circ \leq i \lesssim 30^\circ$) our models predict P~Cygni
profiles with a double peaked absorption lines (see Figure~15).
Such features have been reported in low inclination CVs.
Figure~15 may be compared with the line profiles observed in HR~Del
(\cite{hut80}; \cite{kra81}),
which has an inclination angle of $i \sim 25^\circ$ (\cite{fri92}).

For intermediate and high inclination angles
($55^\circ \lesssim i \leq~90^\circ$)
our model predicts P~Cygni profiles without the absorption component,
which is also consistent with observations
(\cite{kra81}; \cite{hol82}; \cite{cor85}; \cite{mas95}; \cite{che97}).

For intermediate inclination angles
($55^\circ \lesssim i \lesssim 70^\circ$)
our model predicts P~Cygni profiles with
double peaked emission lines (see Figure~17).
Such features have been reported in intermediate inclination CVs.
Figure~17 may be compared with the line profiles observed
in UX~UMa (\cite{hol82}; \cite{mas95}),
which has an inclination angle of $i \sim 65^\circ$ (\cite{fra81}).
Our models suggest that the double peaked \ion{C}{4} emission line
observed in CVs may be interpreted as a wind feature.

For high inclination angles
($70^\circ \lesssim i \lesssim 85^\circ$)
our model predicts P~Cygni profiles with single peaked emission lines
with a FWHM in the order of
$2500 \; {\rm km} \; {\rm s}^{-1}$
(see Figure~18).
Singled peaked emission lines are observed in
high inclination CVs.
Figure~18 may be compared with the line profiles observed in
RW~Tri (\cite{cor85}),
which has an inclination angle of $i \sim 75^\circ$
(\cite{kai83}).

For nearly perpendicular inclination angles
($85^\circ \lesssim i \leq 90^\circ$)
our model predicts P~Cygni profiles with single peaked emission lines,
with a value of the FWHM about $1000 \; {\rm km} \; {\rm s}^{-1}$
less than high inclinations,
presenting two satellite peaks due to emission by material near
the disk surface (see Figure~19).
Figure~19 may be compared with the line profiles observed in
DQ~Her (\cite{cor85}),
which has an inclination angle of $i \sim 90^\circ$
(\cite{cor85}) and which shows a comparable FWHM and two satellite
peaks.

Thus the line profiles for \ion{C}{4} (1550~\AA) obtained through our
computational accretion disk wind model are consistent with
observations in their general form,
in particular in the maximum absorption at roughly half the terminal
velocity for the blue-shifted component,
in the magnitudes of the wind velocities implied by the absorption
components,
in the FWHM of the emission components,
and in the strong dependence in inclination angle.

However,
we wish to emphasize that the models presented here do not self
consistently calculate the wind ionization balance and therefore can
only be considered as a preliminary indication that our models are
consistent with observations.
We plan to present more accurate models in our next paper.

\section{Summary and Conclusions}

We have developed a 2.5-dimensional hydrodynamic line-driven
adiabatic accretion disk wind model.
Our model solves a complete set of adiabatic hydrodynamic partial
differential equations,
using the PPM numerical scheme and implementing the radial temperature
and radiation emission distributions on the surface of an accretion
disk.
From our models we calculate wind mass-loss rates,
terminal velocities,
and line profiles for \ion{C}{4} (1550~\AA) for various angles.
For typical cataclysmic variable parameter values of disk luminosity
$L_{disk}=L_{\sun}$,
white dwarf mass $M_{wd}=0.6M_{\sun}$,
and white dwarf radii $R_{wd}=0.01R_{\sun}$,
we obtain a wind mass-loss rate of
$\dot M_{wind}=8 \times 10^{-12} M_{\sun} {\rm yr}^{-1}$,
and a terminal velocity of $\sim 3000 {\rm \;km \;s}^{-1}$.
Studies of observed ultraviolet line profiles from CVs have estimated
CV  wind mass loss rates to be in the order of
$10^{-12}$ to $10^{-11} M_{\sun} {\rm yr}^{-1}$ and found CV wind
terminal speeds in the order of $\sim 3000 \; {\rm km} \; {\rm s}^{-1}$
(\cite{kra81}; \cite{cor82}; \cite{gre82}; \cite{mau87}; \cite{pri95};
\cite{fri97}),
thus the models developed in this work predict values for
wind mass loss rates and terminal speed consistent with observations.

Furthermore,
the line profiles we obtain through our computational model
are also consistent with observations in their general form,
in particular in the maximum absorption at roughly half the terminal
velocity for the blue-shifted component,
in the magnitudes of the wind velocities implied by the absorption
components,
in the FWHM of the emission components,
and in the strong dependence in inclination angle.

We have shown that rotational forces are important in the study of
winds from accretion disks.
They cause the velocity streamlines to collide which results in
an enhanced density region.
In this region the wind speed is reduced and the wind density is
increased.
The increase in density caused by the collision of streamlines is
important because it permits the appearance of blue-shifted absorption
lines as observed in P~Cygni profiles of low-inclination CVs.

A shortcoming of our models is the neglect of radiative cooling
and heating and local ionization equilibrium throughout the wind.
In the next paper of this series we will present models which include
these effects.
We also plan to extend our models to the study of the wind dynamics
of low mass X-ray binaries and active galactic nuclei,
where these processes may play a more important role in determining the
overall dynamics.

\acknowledgments

We thank J. Stone for useful comments and suggestions,
and the referee for many constructive suggestions on the
manuscript.

\clearpage

\figcaption[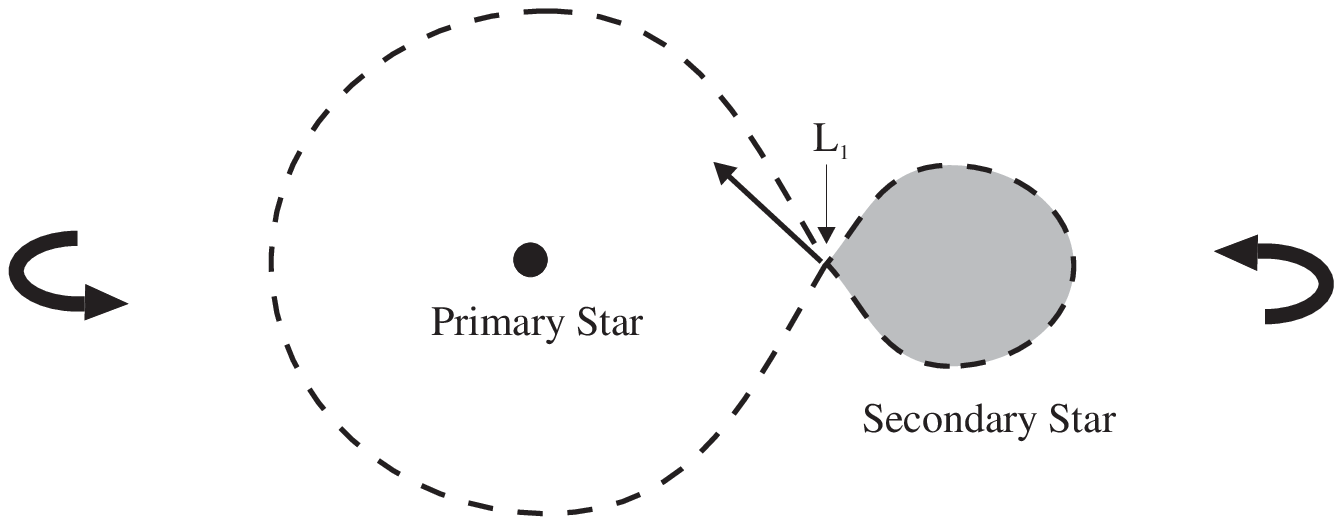]{Cataclysmic variable system: the secondary star
fills it Roche lobe (dotted line) and transfers mass to the white
dwarf or primary star through the inner Lagrangian point (L$_1$),
while both stars rotate about their respective center of mass.}

\figcaption[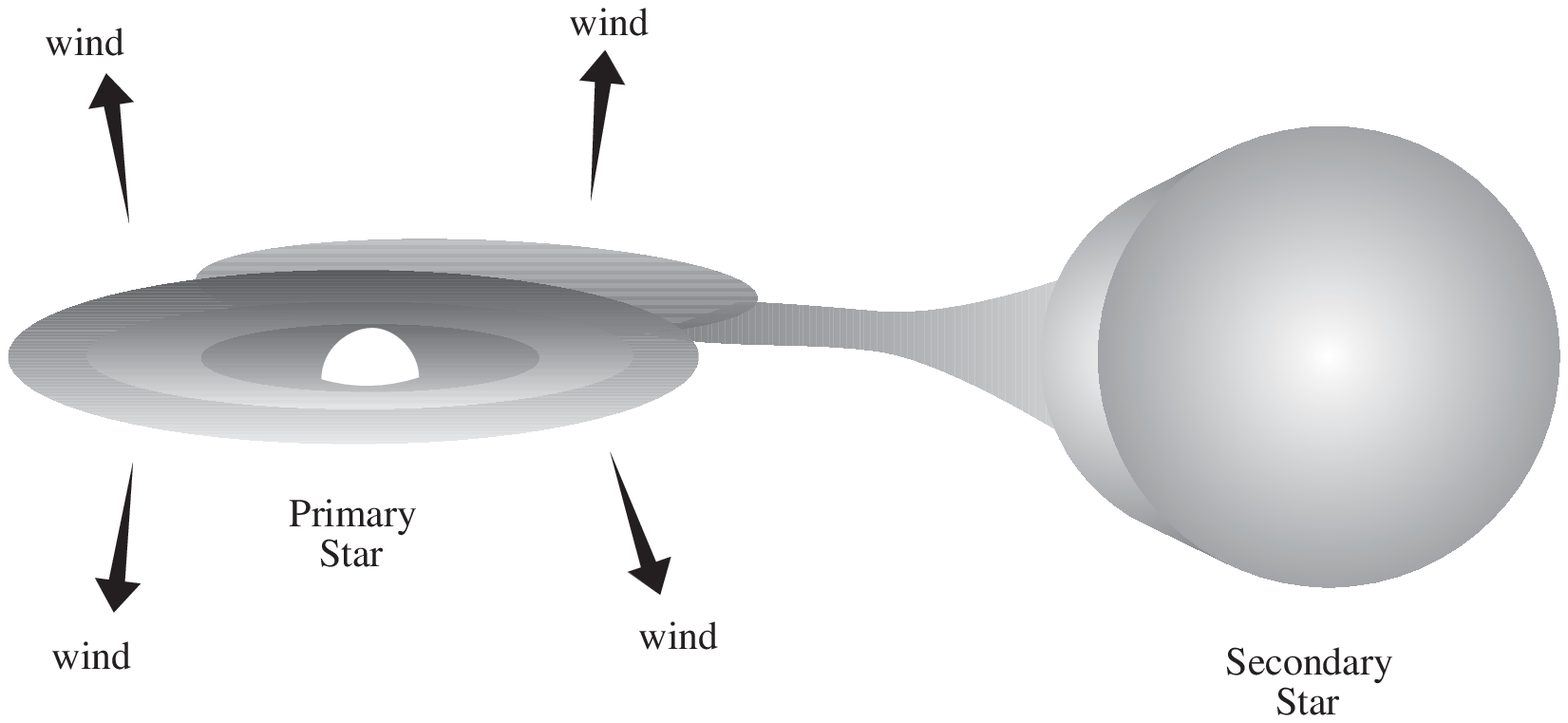]{Accretion disk winds in cataclysmic variables}

\figcaption[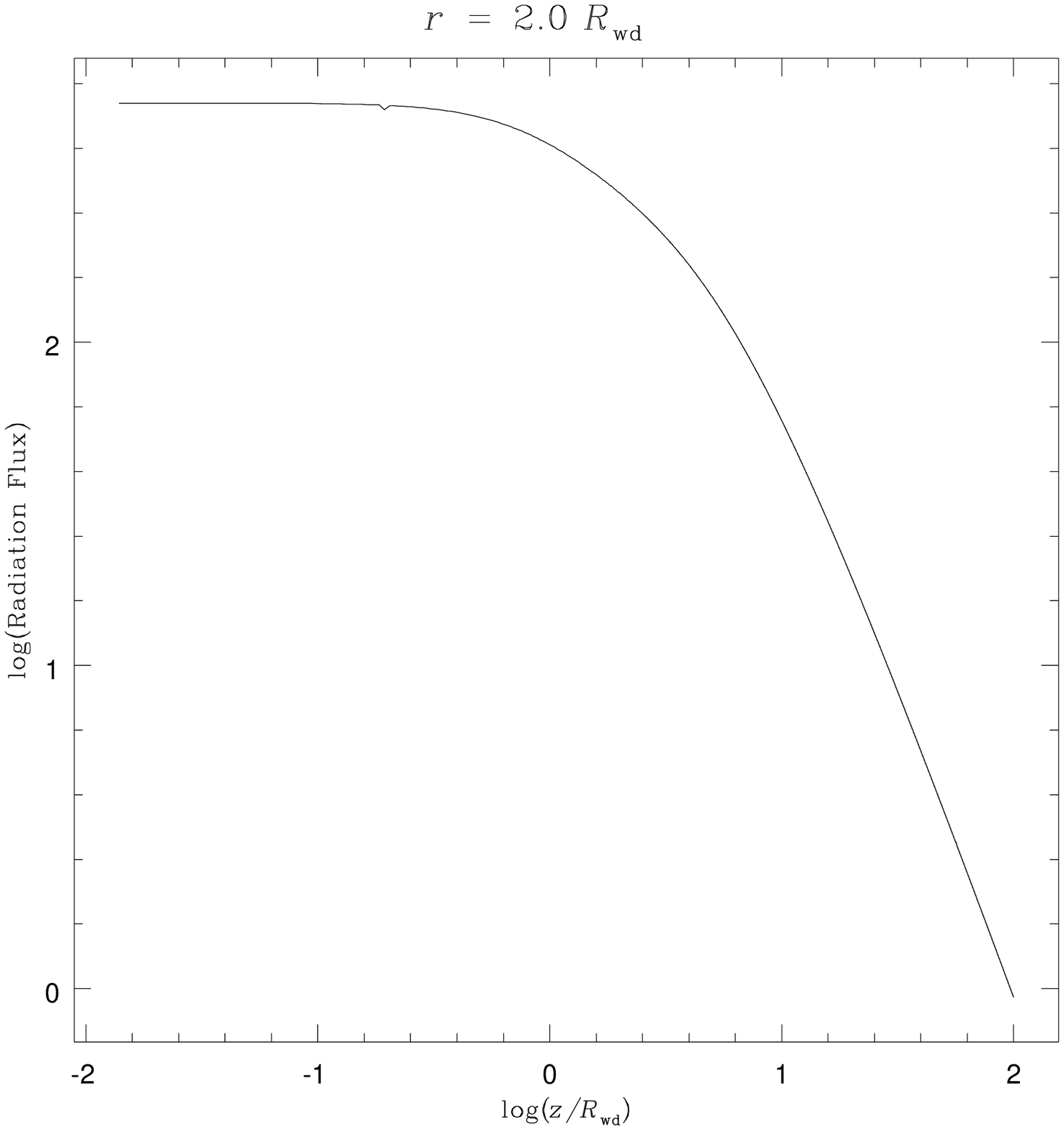,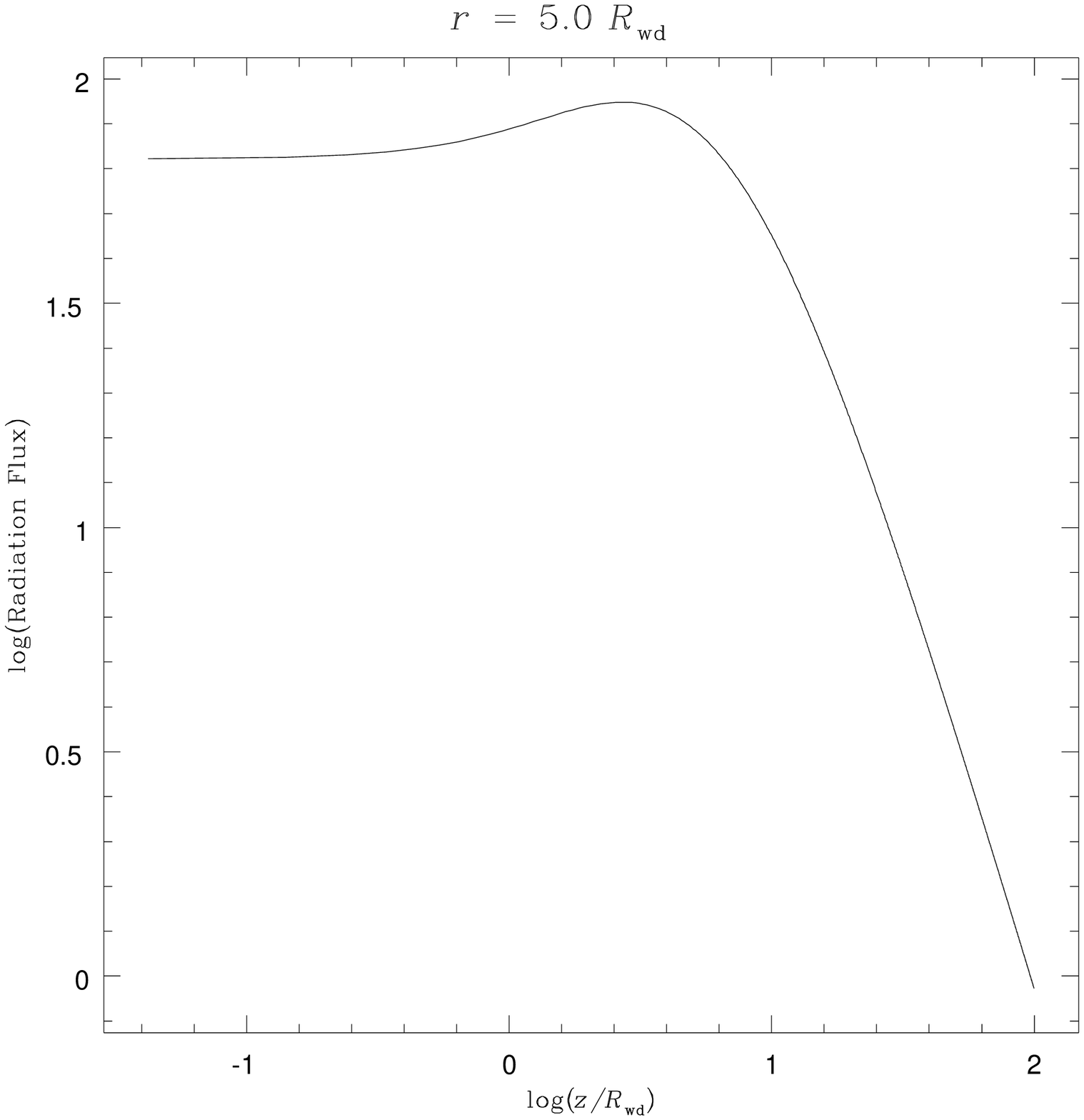,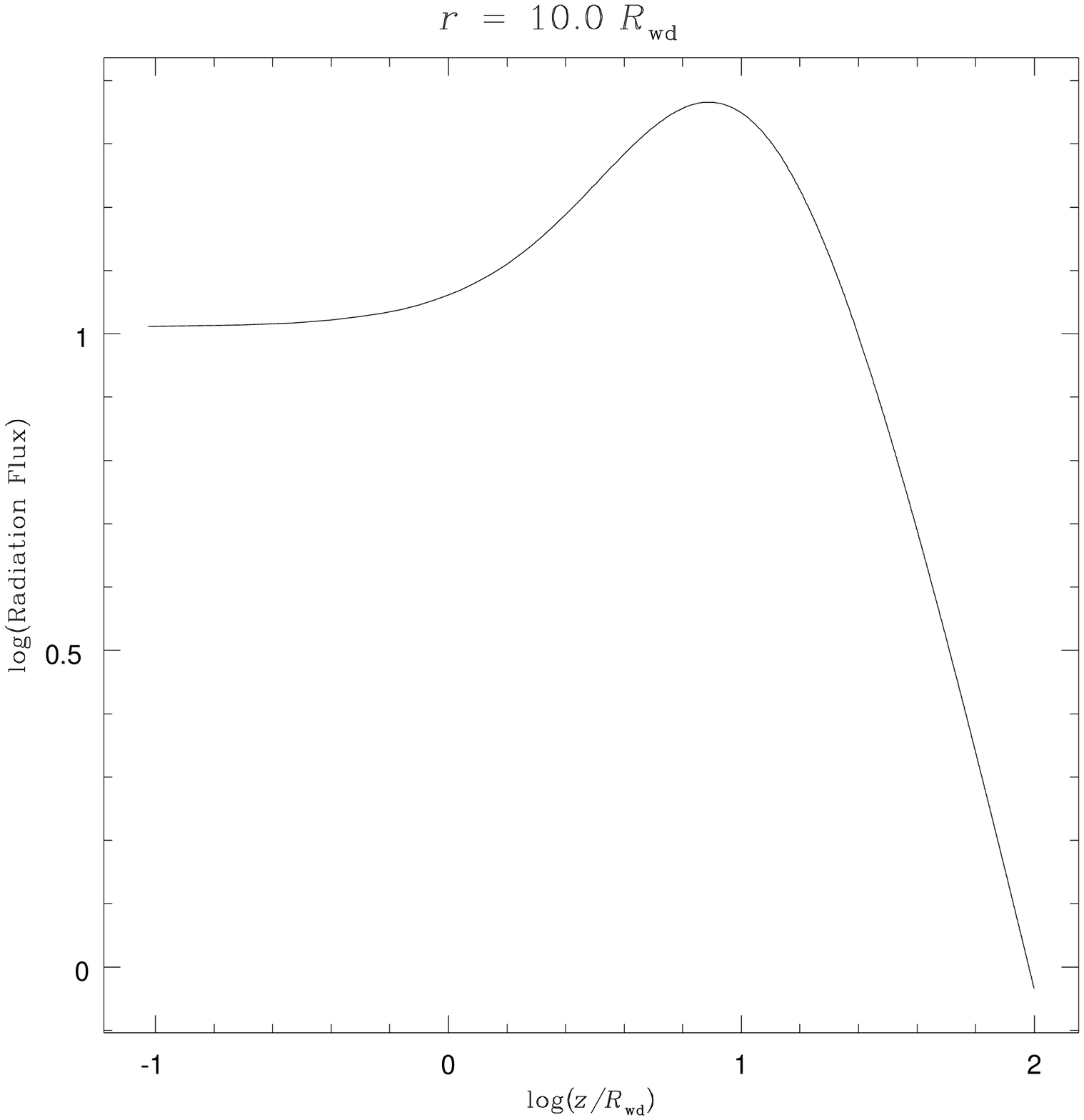,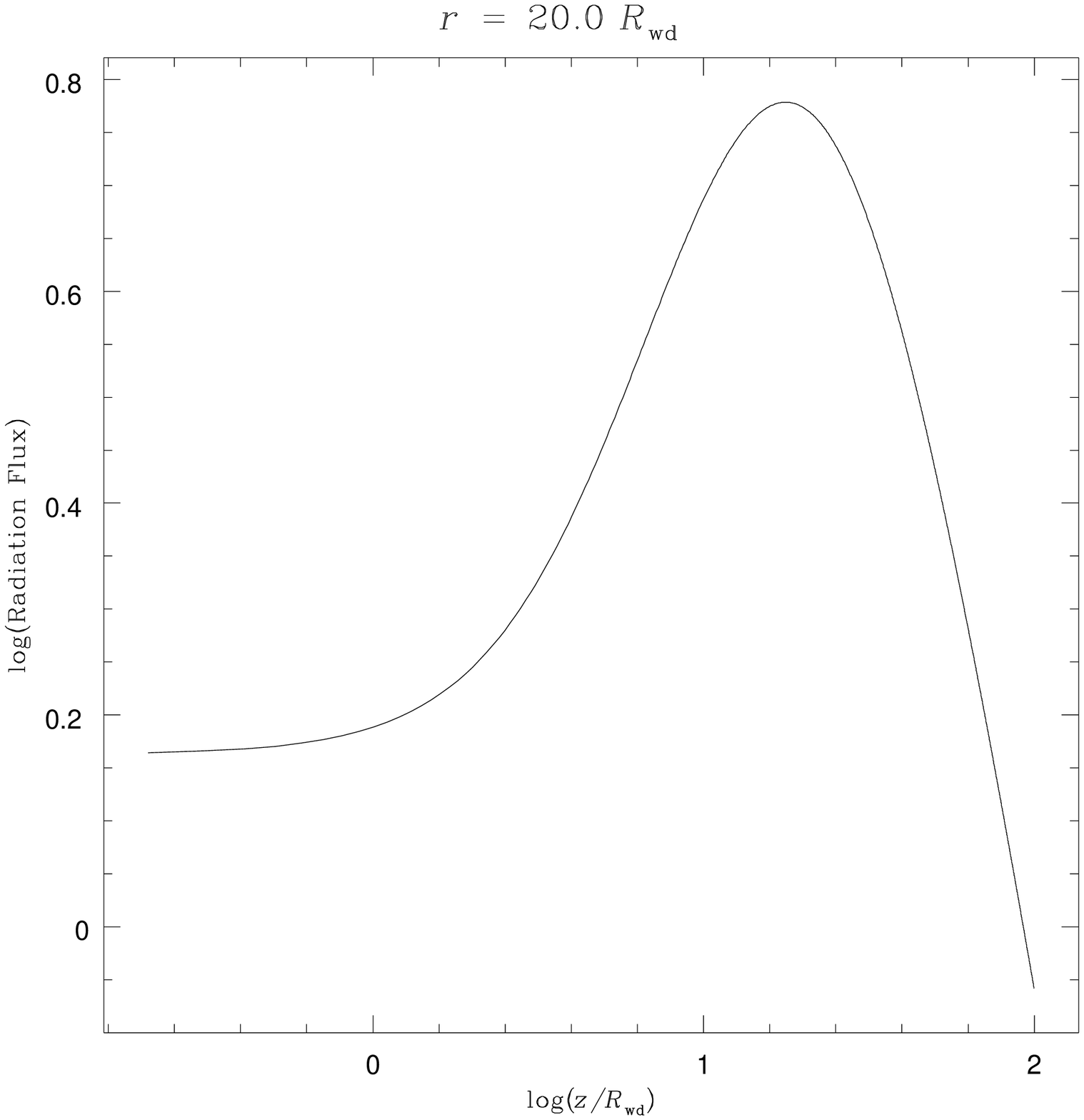,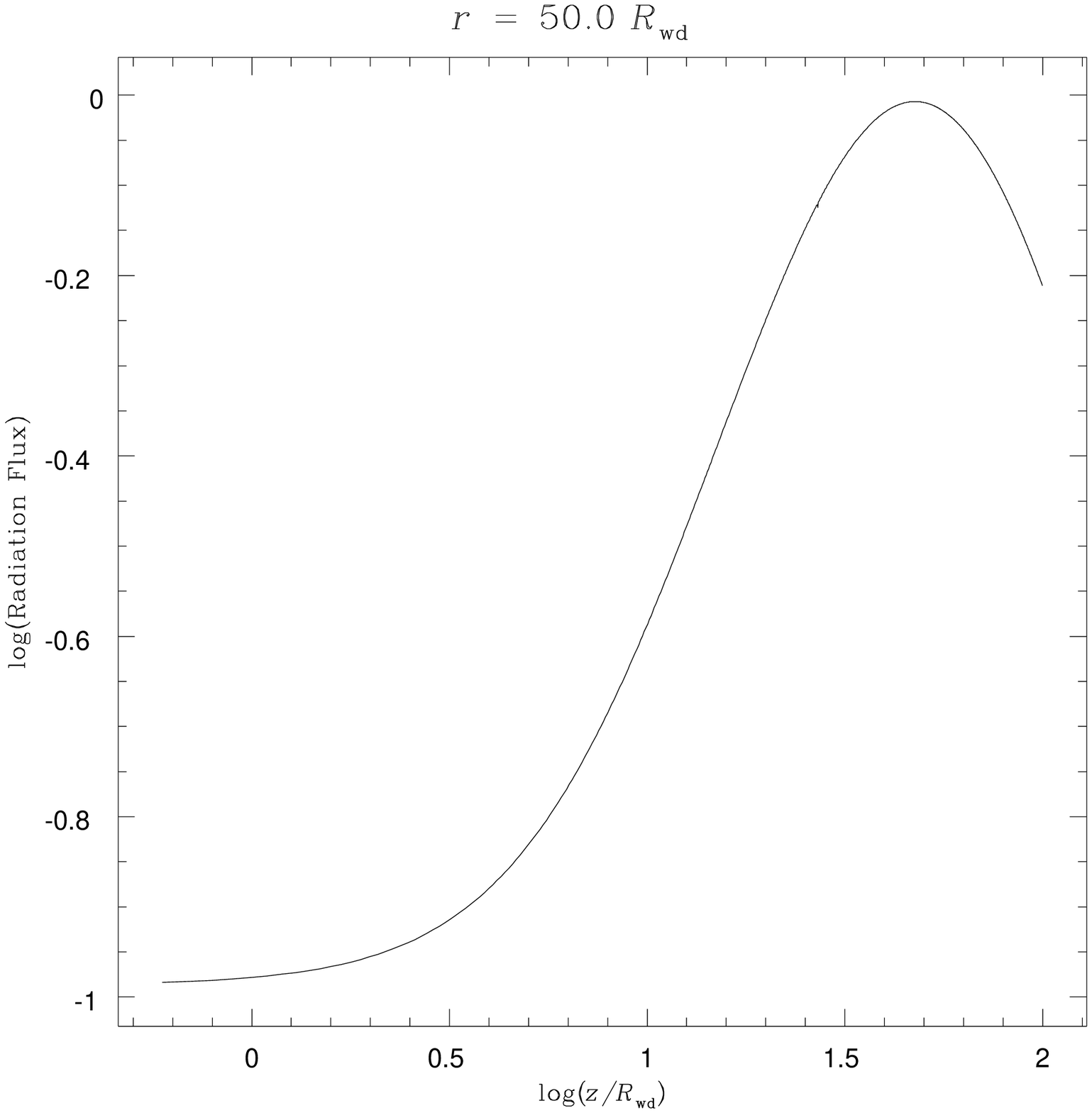,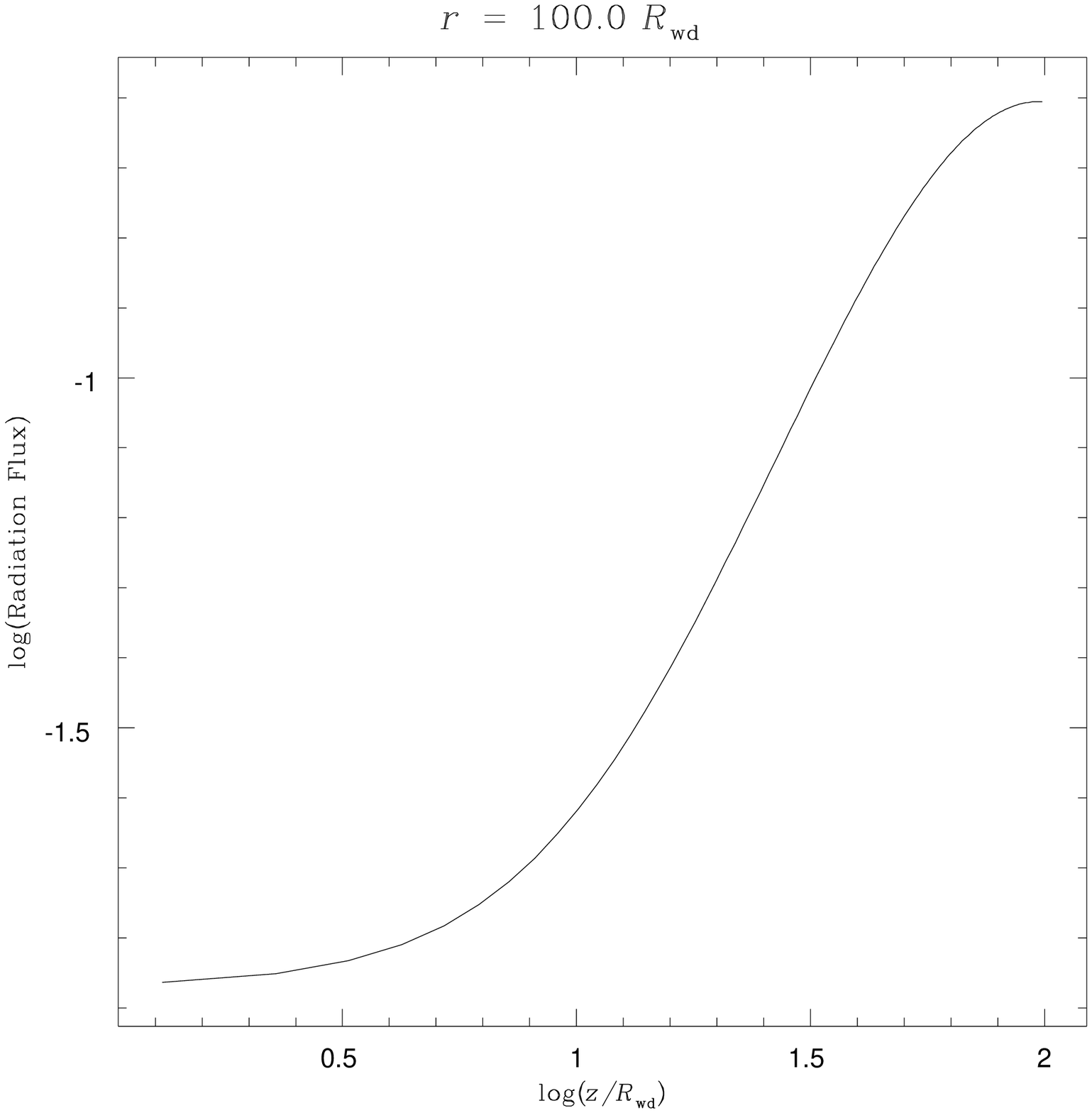]
{Logarithm of the radiation flux perpendicular to the disk in units of
$L_{\sun} / 2 R_{\sun}^2$ vs. the logarithm of the height above the
disk in units of white dwarf radii $R_{wd}$ for several radii.
The physical parameters used here are $L_{disk}=L_{\sun}$ and
$R_{wd} = 0.01 R_{\sun}$.
({\it a})~$r=2R_{wd}$~;
({\it b})~$r=5R_{wd}$~;
({\it c})~$r=10R_{wd}$~;
({\it d})~$r=20R_{wd}$~;
({\it e})~$r=50R_{wd}$~;
({\it f})~$r=100R_{wd}$~. }

\figcaption[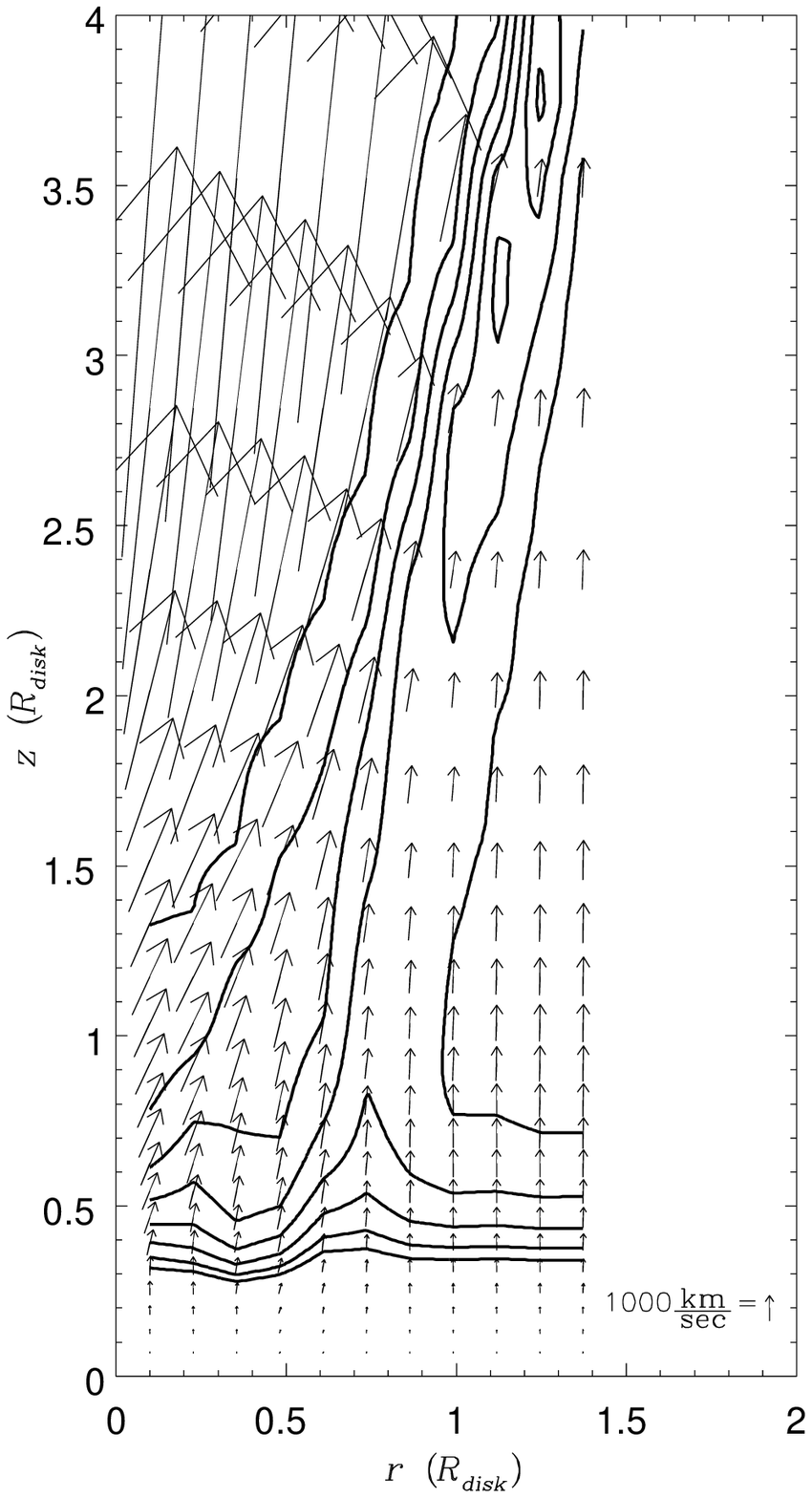]
{Vector field graph of wind velocity superimposed
with a contour graph of wind density for the two-dimensional
isothermal disk wind model
(Paper~1: isothermal disk and isothermal wind).
The primary star is at the origin of the graph,
and the disk is over the horizontal axis.
The contour levels vary uniformly from a value of
$4.2 \times 10^{-19} \; {\rm g \; cm}^{-3}$ down to a value of
$0.1 \times 10^{-19} \; {\rm g \; cm}^{-3}$.
The physical parameters here used are $M_{wd} = 0.6 M_{\sun}$,
$R_{wd}=0.01R_{\sun}$,
and $L_{disk} = L_{\sun}$.}

\figcaption[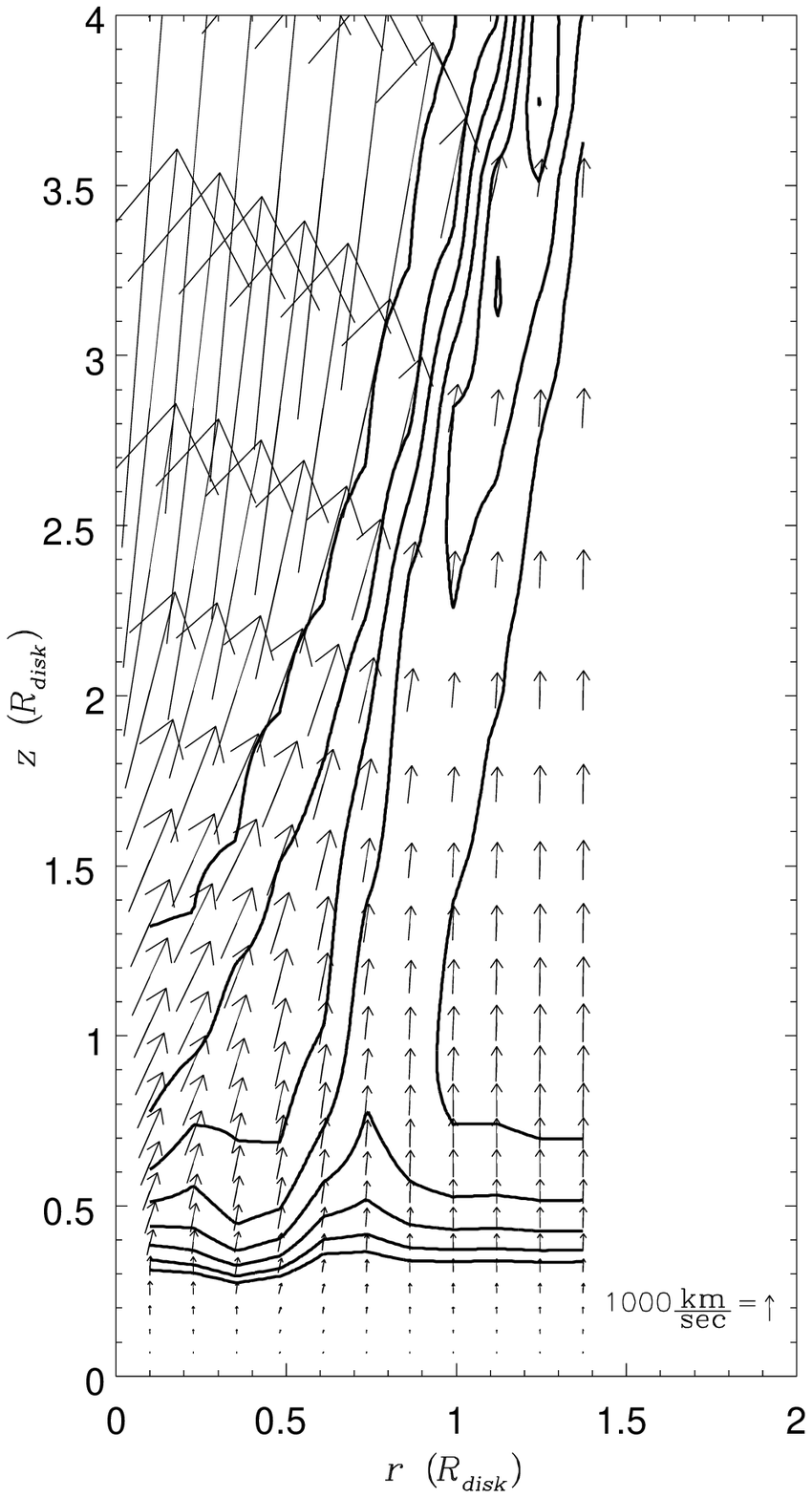]
{Vector field graph of wind velocity superimposed
with a contour graph of wind density for the two-dimensional
isothermal disk wind model with values presented by
Abbott~(1982) implemented for the force multiplier parameters
(rather than the earlier values used by Castor, Abbott, \& Klein~(1975)
as we implemented in the models presented in Paper~1).
Note the higher density values found in this model with respect to
the model represented in Figure~4~.
The primary star is at the origin of the graph,
and the disk is over the horizontal axis.
The contour levels vary uniformly from a value of
$11.34 \times 10^{-18} \; {\rm g \; cm}^{-3}$ down to a value of
$0.27 \times 10^{-18} \; {\rm g \; cm}^{-3}$.
The physical parameters here used are $M_{wd} = 0.6 M_{\sun}$,
$R_{wd}=0.01R_{\sun}$,
and $L_{disk} = L_{\sun}$.}

\figcaption[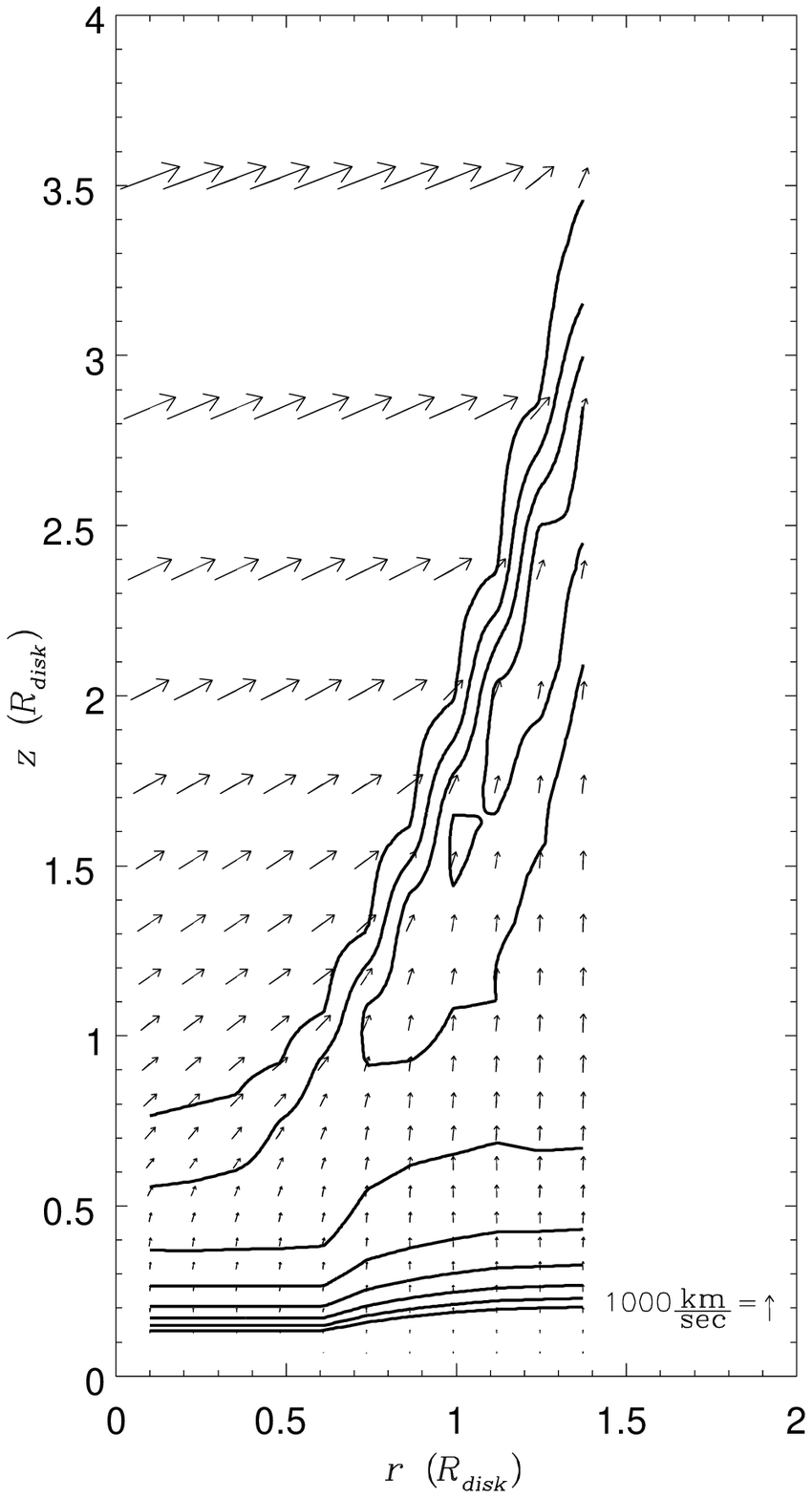]
{Vector field graph of wind velocity superimposed
with a contour graph of wind density for the two-dimensional
isothermal disk wind model implementing the fact that the radiation
pressure due to lines arrives at a maximum value for sufficiently high
velocity gradients or low density values.
Note the higher density values found in this model with respect to
the model represented in Figure~5~.
The primary star is at the origin of the graph,
and the disk is over the horizontal axis.
The contour levels vary uniformly from a value of
$15.0 \times 10^{-17} \; {\rm g \; cm}^{-3}$ down to a value of
$1.0 \times 10^{-17} \; {\rm g \; cm}^{-3}$.
The physical parameters here used are the same as in Figure~5~.}

\figcaption[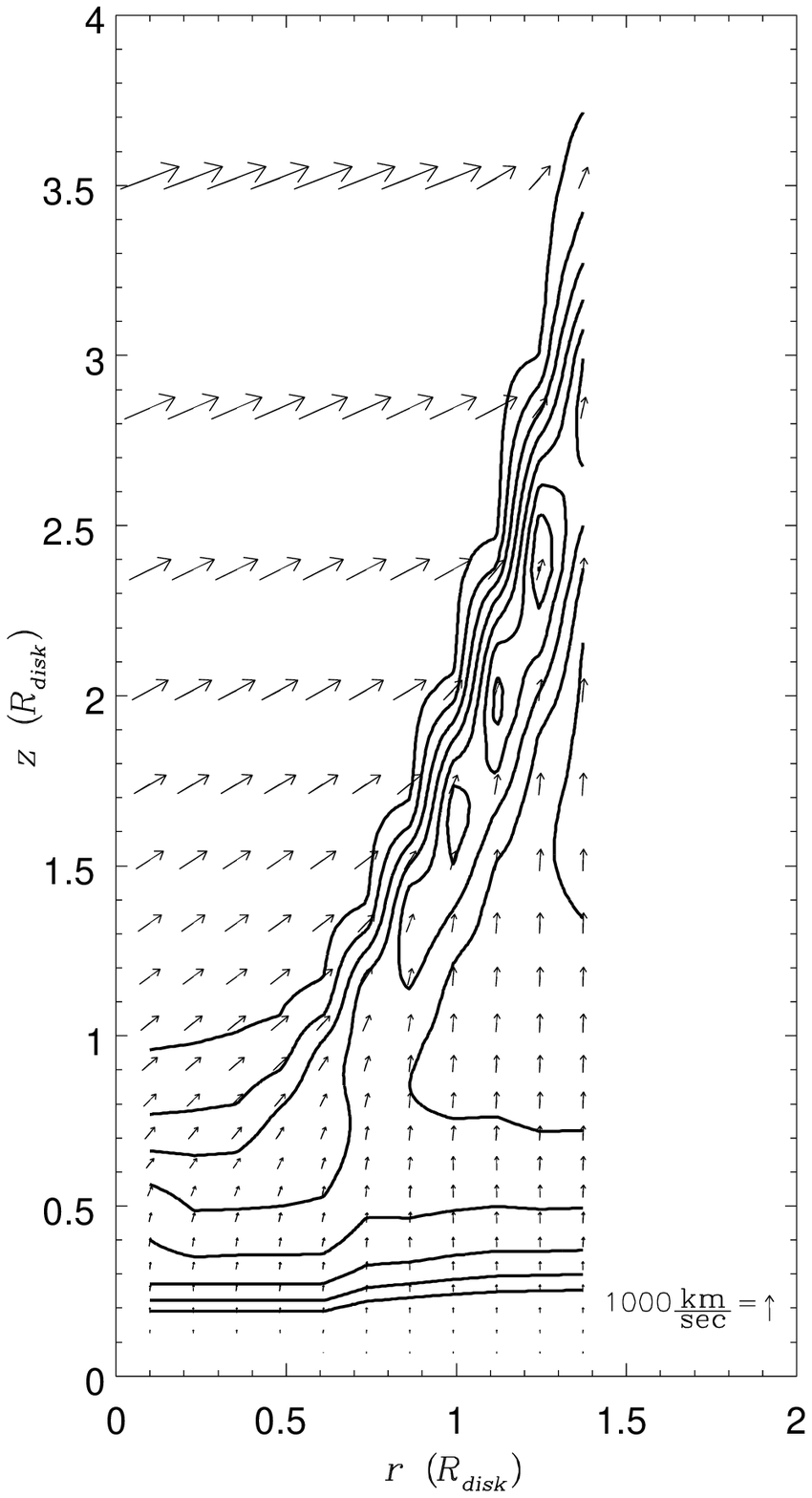]
{Vector field graph of wind velocity superimposed
with a contour graph of wind density for the two-dimensional
adiabatic wind isothermal disk model.
The primary star is at the origin of the graph,
and the disk is over the horizontal axis.
The contour levels vary uniformly from a value of
$15.0 \times 10^{-17} \; {\rm g \; cm}^{-3}$ down to a value of
$1.0 \times 10^{-17} \; {\rm g \; cm}^{-3}$.
The physical parameters here used are the same as in Figure~6~.}

\figcaption[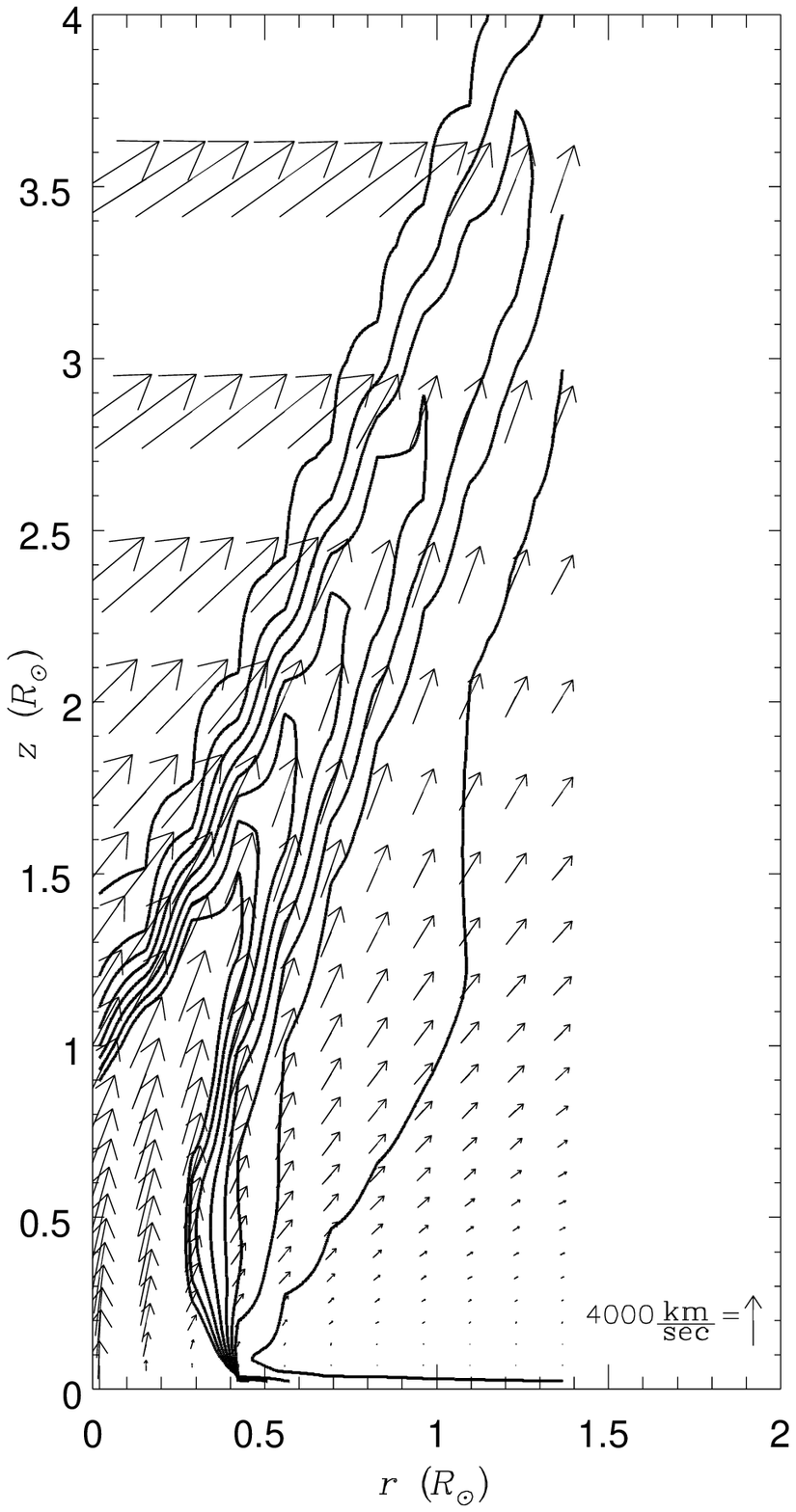]
{Vector field graph of wind velocity superimposed
with a contour graph of wind density for the two-dimensional
accretion disk wind model.
The primary star is at the origin of the graph,
and the disk is over the horizontal axis.
The contour levels vary uniformly from a value of
$3.6 \times 10^{-16} \; {\rm g \; cm}^{-3}$ down to a value of
$0.1 \times 10^{-16} \; {\rm g \; cm}^{-3}$.
The physical parameters here used are $M_{wd} = 0.6 M_{\sun}$,
$R_{wd}=0.01R_{\sun}$,
and $L_{disk} = L_{\sun}$.}

\figcaption[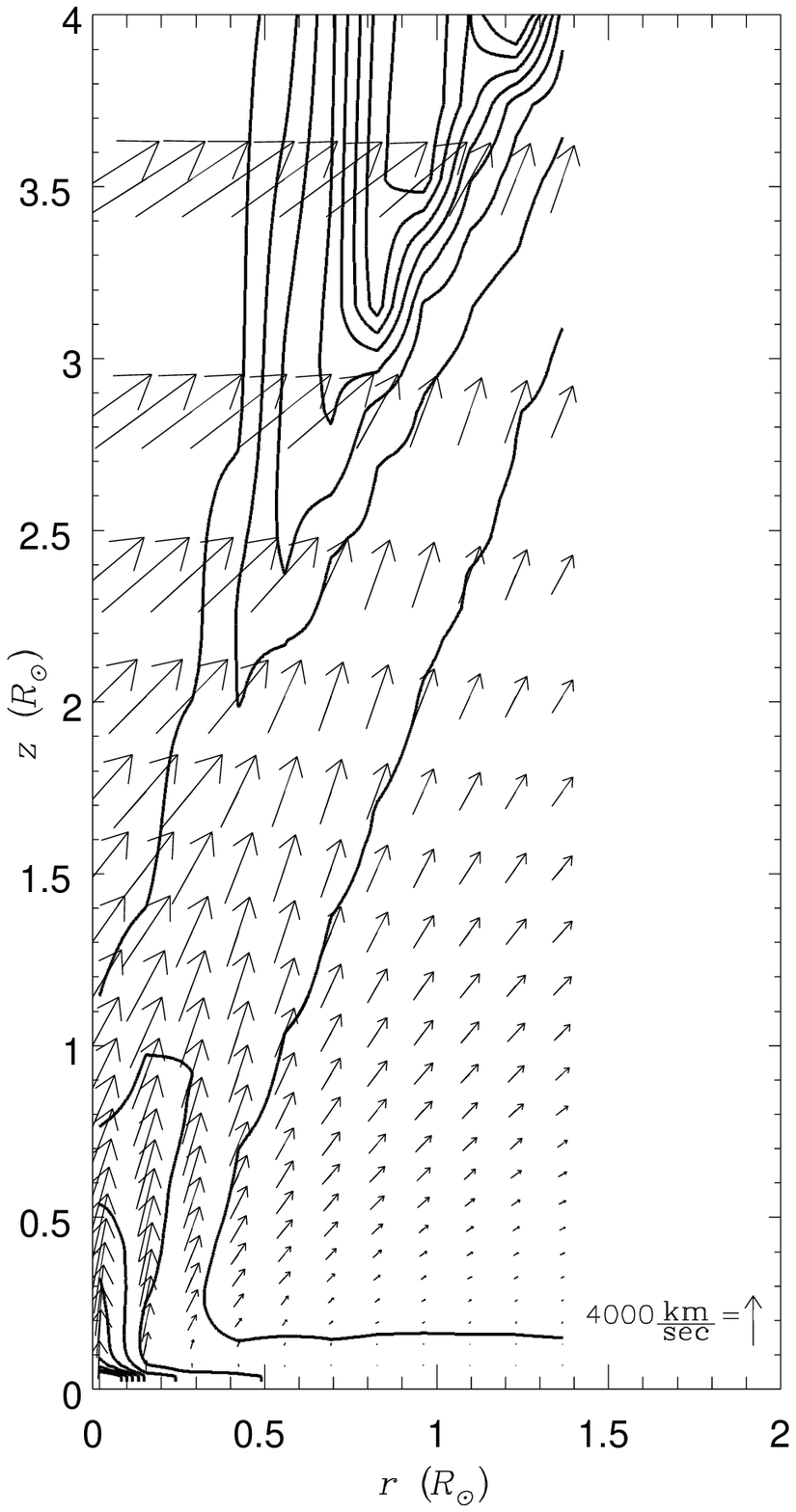]
{Vector field graph of wind velocity superimposed
with a contour graph of wind temperature for the two-dimensional
accretion disk wind model.
The primary star is at the origin of the graph,
and the disk is over the horizontal axis.
The contour levels vary uniformly from a value of
$22,000 \; {\rm K}$  down to a value of $1,000 \; {\rm K}$.
The physical parameters here used are the same as in Figure~8~.}

\figcaption[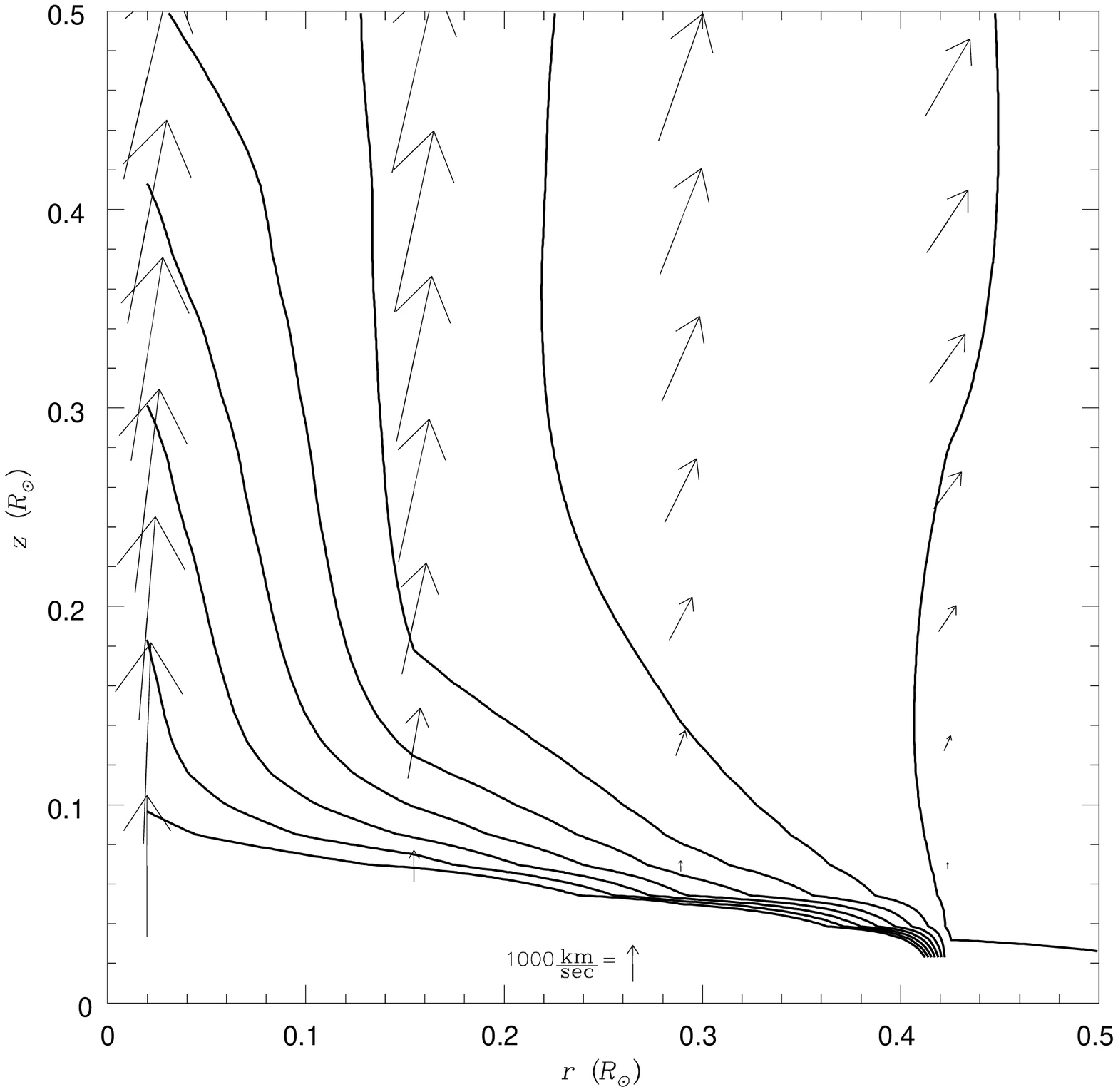,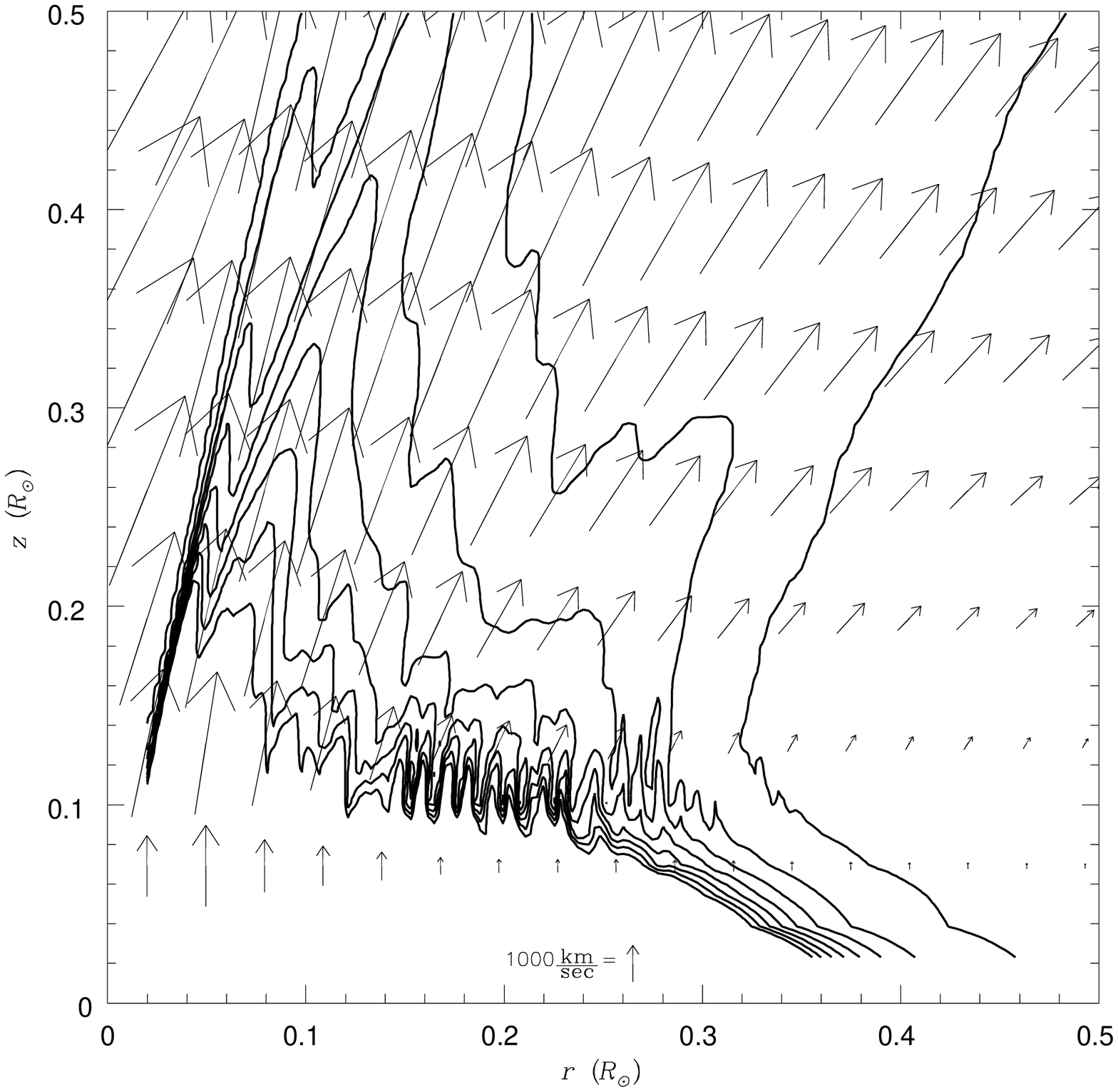]
{Vector field graph of wind velocity superimposed
with a contour graph of wind density for the two-dimensional accretion
disk wind model for ({\it a}) 11 grid points in the $r$ direction;
({\it b}) $1001$ grid points in the $r$ direction.
The primary star is at the origin of the each graph,
and the disk is over the horizontal axis of each graph.
The contour levels vary uniformly from a value of
$3.6 \times 10^{-15} \; {\rm g \; cm}^{-3}$ down to a value of
$0.1 \times 10^{-15} \; {\rm g \; cm}^{-3}$.
The physical parameters here used are the same as in Figure~8~.}

\figcaption[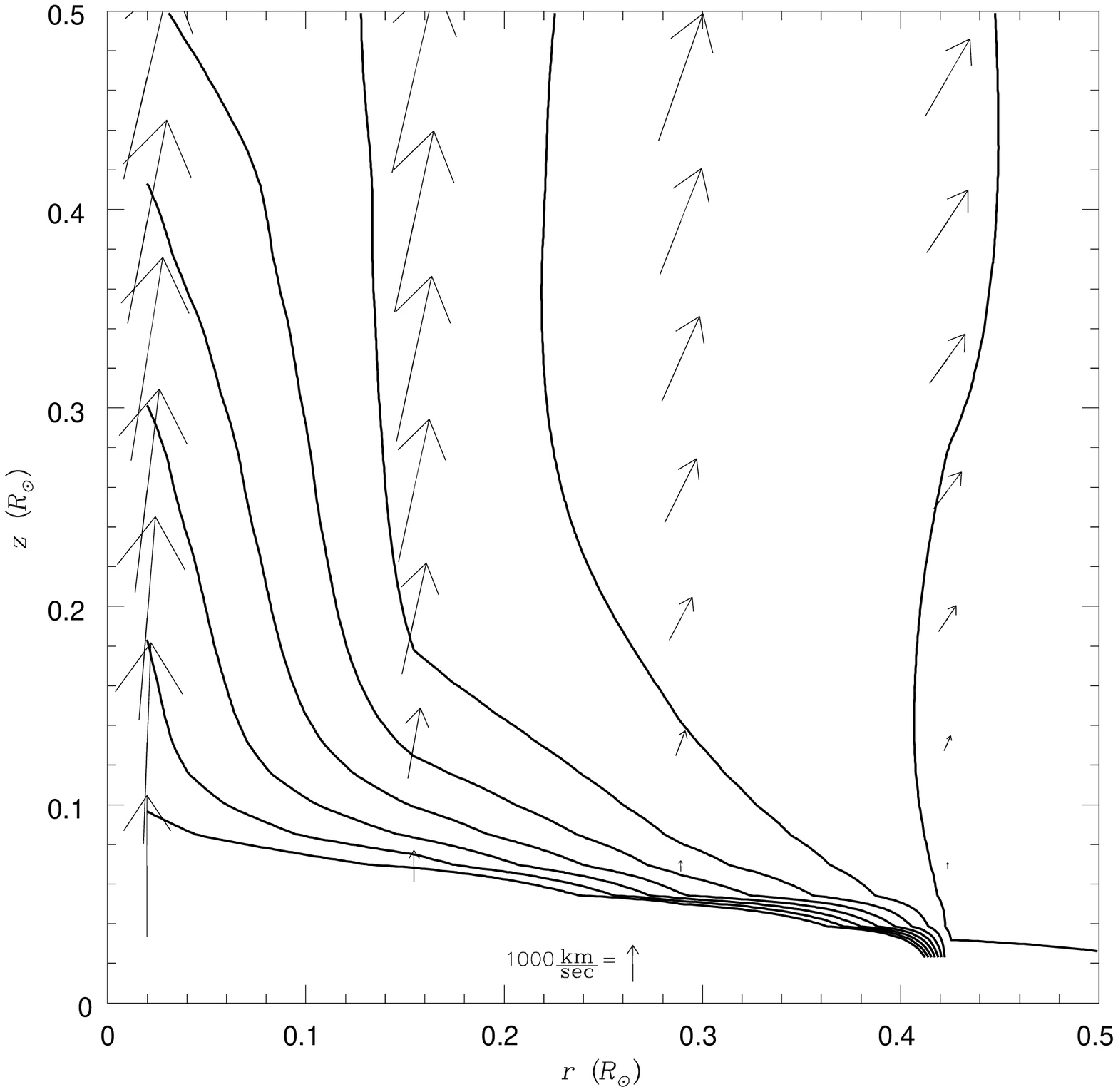,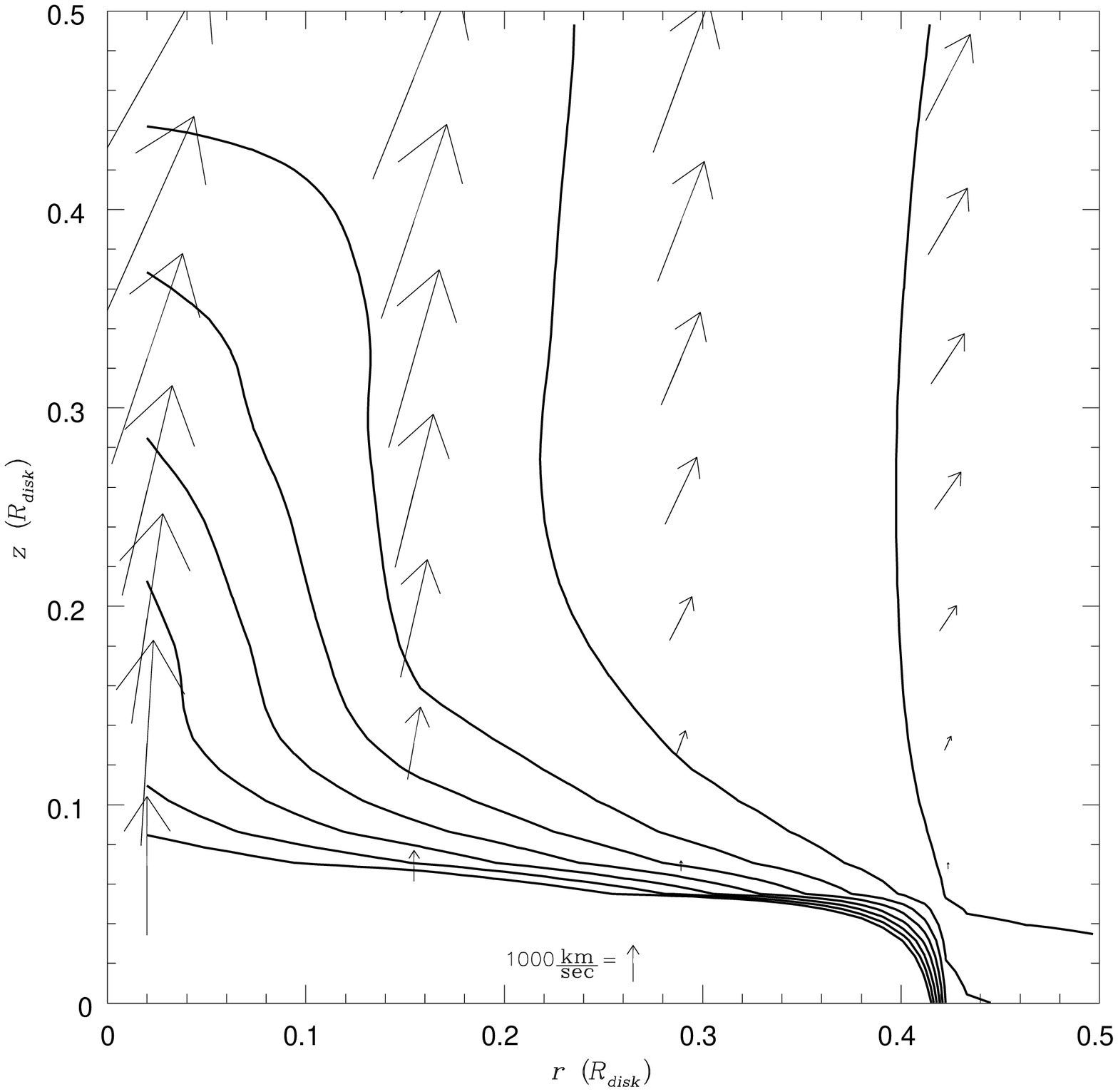]
{Vector field graph of wind velocity superimposed
with a contour graph of wind density for the two-dimensional accretion
disk wind model for
({\it a}) a lower boundary for the computational grid of
$z_0=0.023R_{\sun}$;
({\it b}) a lower boundary for the computational grid of
$z_0=0.00023R_{\sun}$.
The primary star is at the origin of the each graph,
and the disk is over the horizontal axis of each graph.
The contour levels vary uniformly from a value of
$3.6 \times 10^{-15} \; {\rm g \; cm}^{-3}$ down to a value of
$0.1 \times 10^{-15} \; {\rm g \; cm}^{-3}$.
The physical parameters here used are the same as in Figure~8~.}

\figcaption[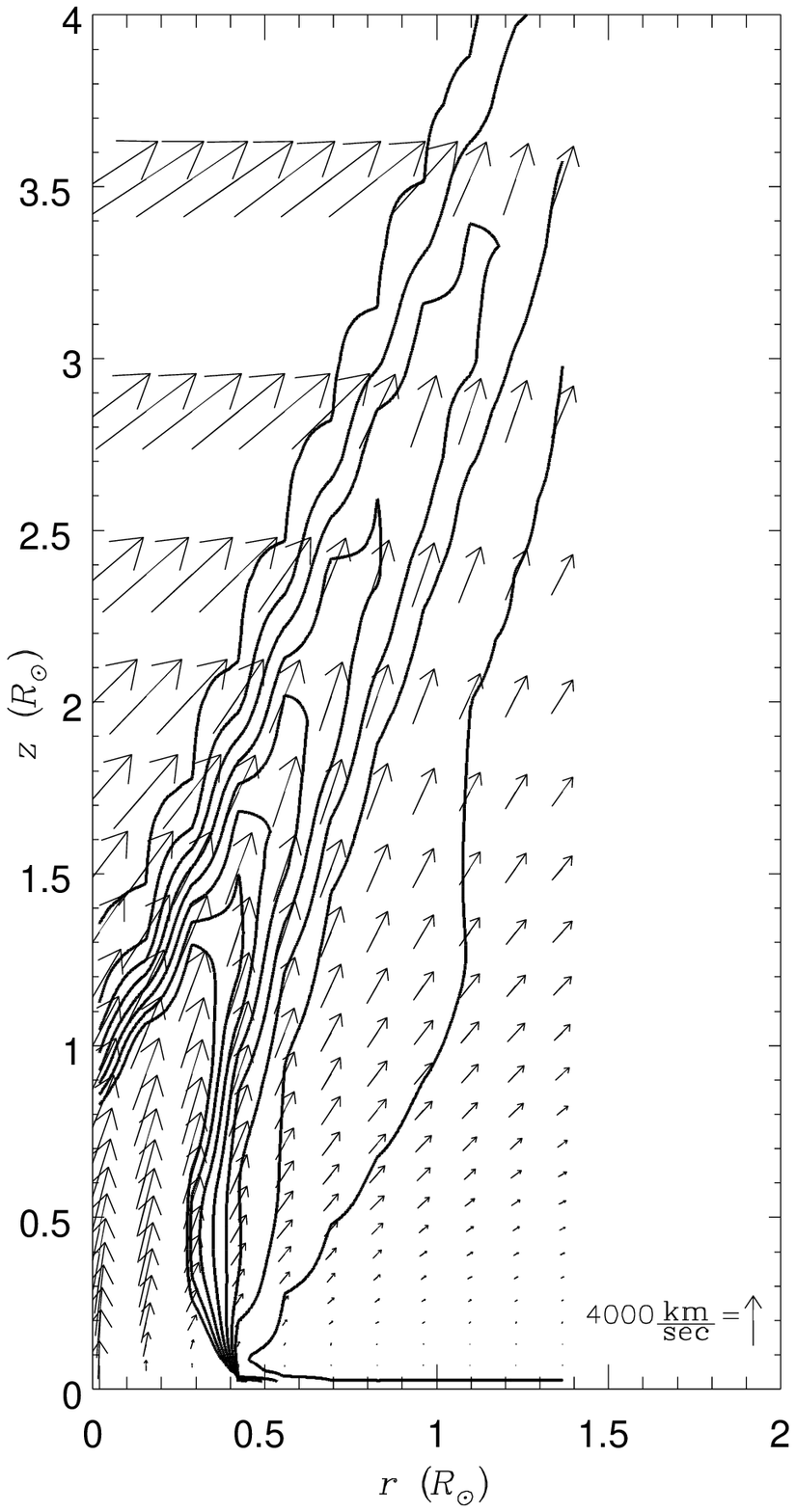]
{Vector field graph of wind velocity superimposed
with a contour graph of wind density for the two-dimensional accretion
disk wind model with a reflecting wall boundary condition on the
left side of the computational grid
(rather than a free boundary condition as in the results presented in
Figure~8).
The primary star is at the origin of the graph,
and the disk is over the horizontal axis.
The contour levels and the physical parameters here used are the same
as in Figure~8~.}

\figcaption[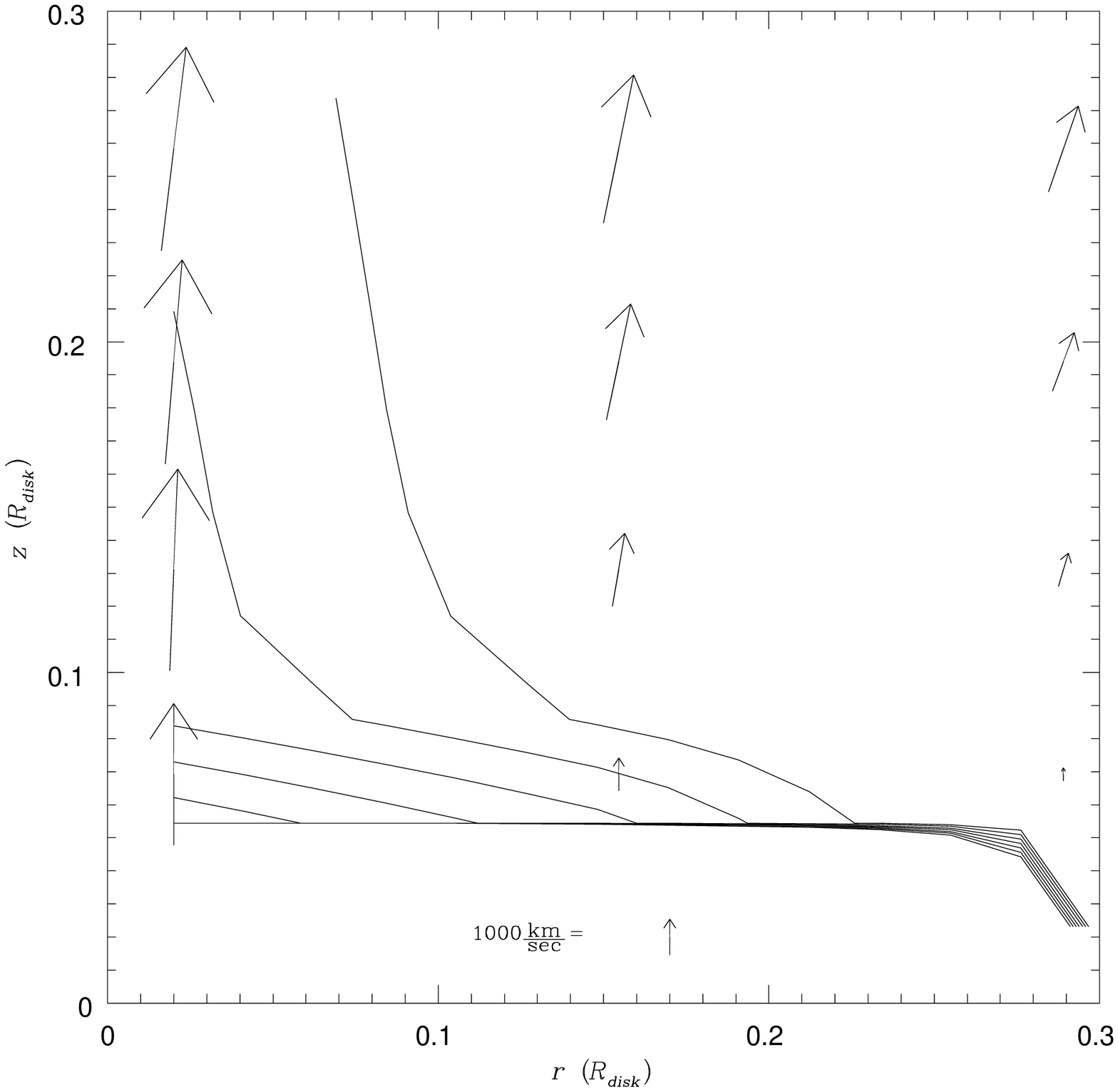,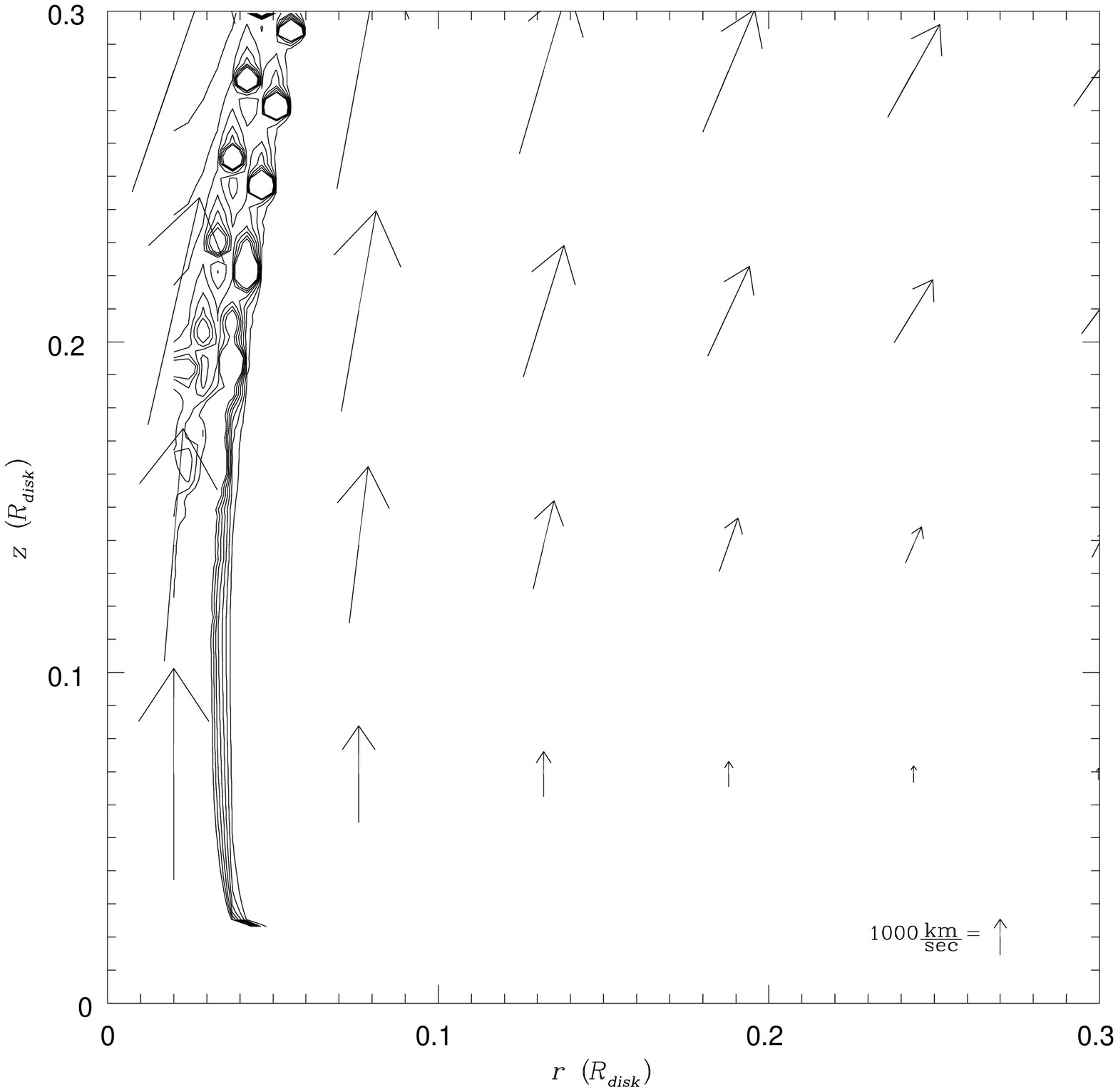,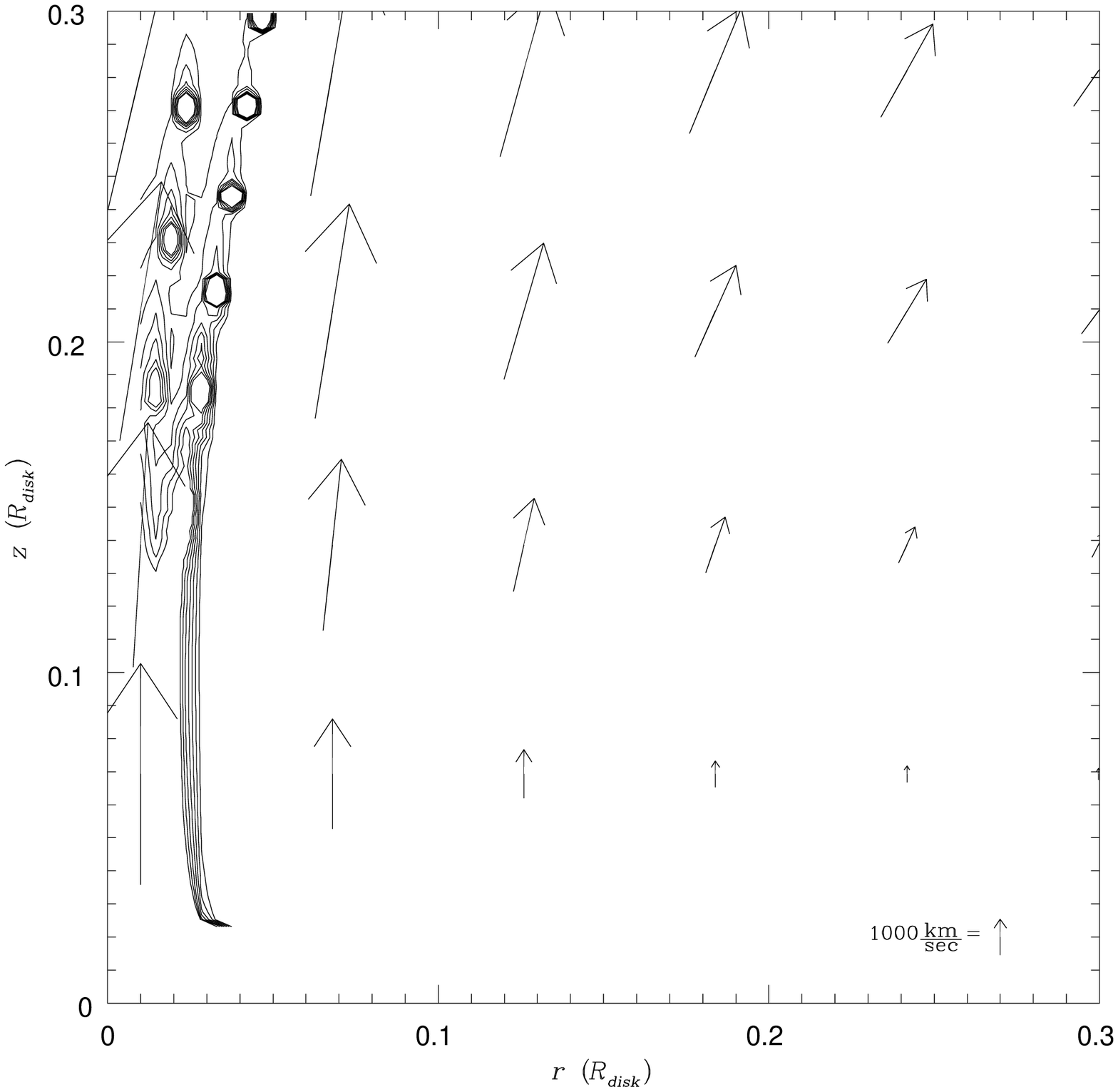]
{Vector field graph of wind velocity superimposed with a contour graph
of wind density for the two-dimensional accretion disk wind model for
({\it a}) $11$ grid points in the $r$ direction, ranging from $0.02 R_{\sun}$
($2.0 R_{wd}$) to $61.0 R_{\sun}$,  and 101 grid points in the $z$ direction
ranging from $0.023R_{\sun}$ to $61.0 R_{\sun}$ (we note here that in this
figure, for both directions, we present only results up to $0.3R_{\sun}$
so as to present the three figures at the same scale);
({\it b}) $1001$ grid points in the $r$ direction, ranging from
$0.02 R_{\sun}$ ($2.0 R_{wd}$) to $0.3 R_{\sun}$,
and $501$ grid points in the $z$ direction, ranging from $0.023 R_{\sun}$
to $0.3 R_{\sun}$;
({\it c}) $1001$ grid points in the $r$ direction, ranging from
$0.01 R_{\sun}$ ($1.0 R_{wd}$) to $0.3 R_{\sun}$,
and $501$ grid points in the $z$ direction, ranging from $0.023 R_{\sun}$
to $0.3 R_{\sun}$.
The primary star is at the origen of the graph, and the disk is over the
horizontal axis.
The contour levels vary uniformly from a value of
$8.0 \times 10^{-15} \; {\rm g \; cm}^{-3}$ down to a value of
$2.0 \times 10^{-15} \; {\rm g \; cm}^{-3}$.
The physical parameters here used are the same as in Figure~8~.}

\figcaption[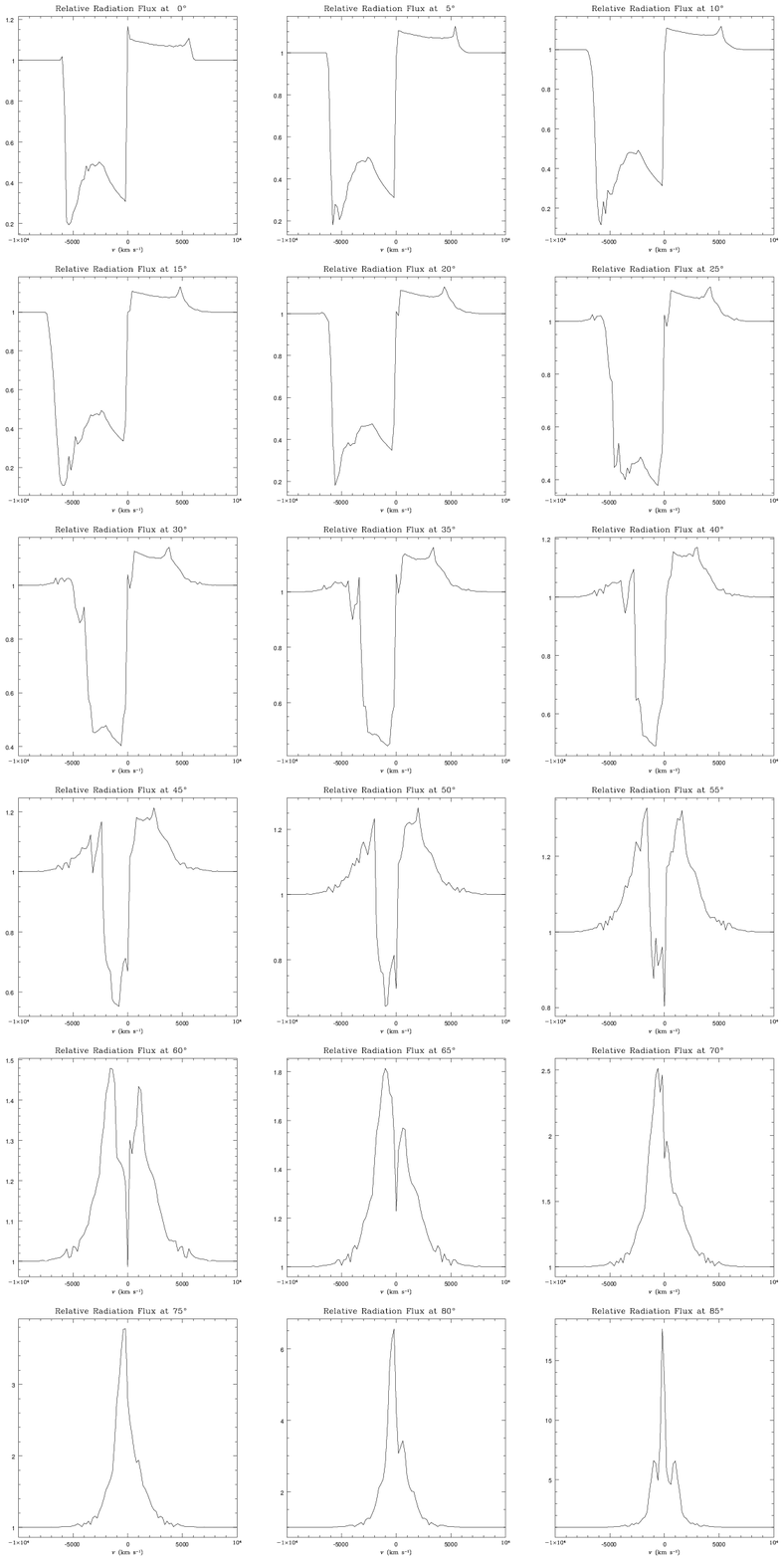]
{\ion{C}{4} 1550~\AA \ line profiles obtained by the two-dimensional
accretion disk wind model for several angles.
The physical parameters here used are $M_{wd} = 0.6 M_{\sun}$,
$R_{wd}=0.01R_{\sun}$,
and $L_{disk} = L_{\sun}$.
In the calculation of these line profiles we have assumed single
scattering and a relative abundance of carbon with respect to
hydrogen of $n_{\rm CIV} / n_{\rm H} = 10^{-3}$ throughout the wind.
The inclinations angles vary from $0^\circ$ to $85^\circ$ in $5^\circ$ 
increments
(left to right and top to bottom).}

\figcaption[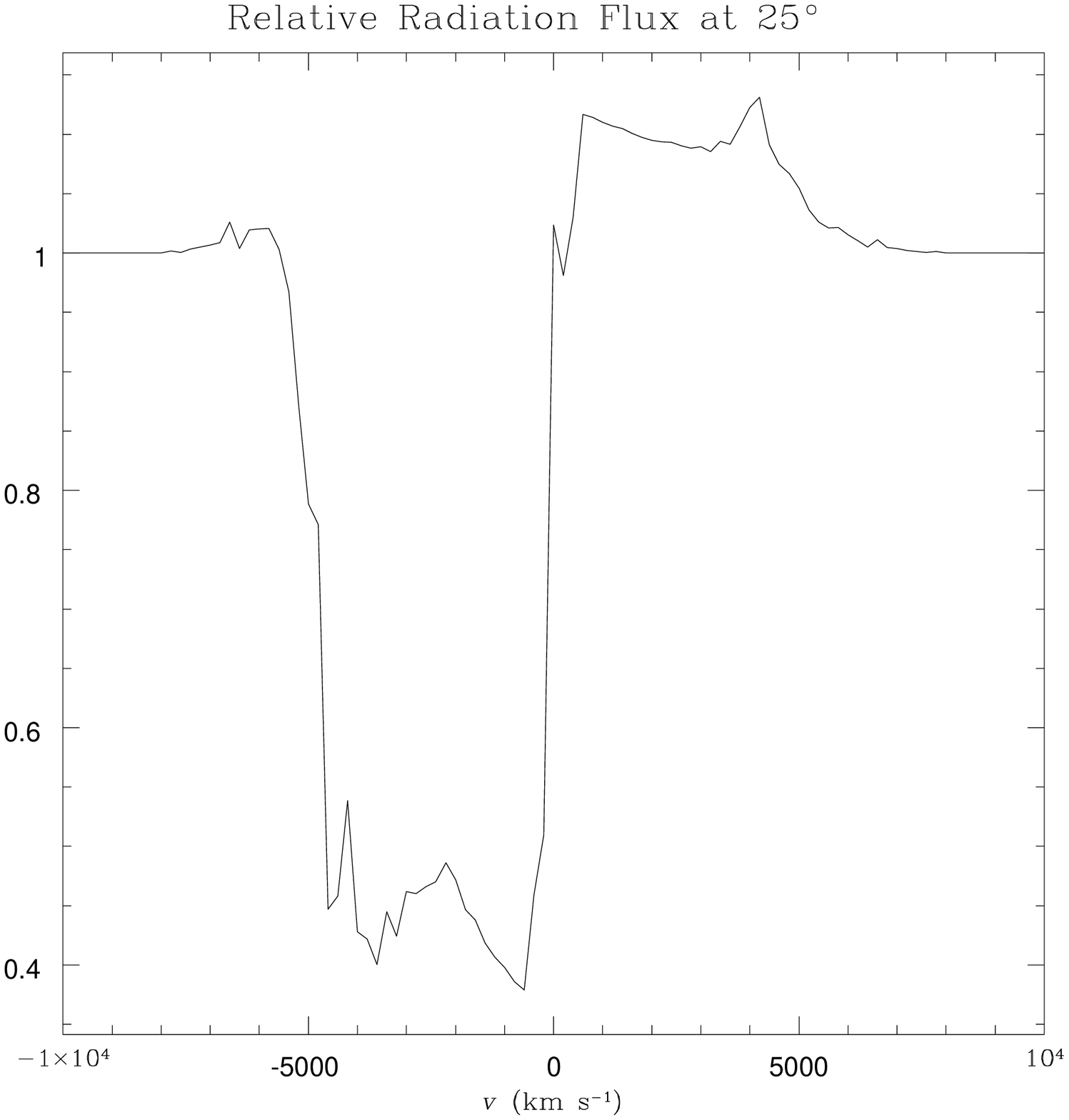]
{\ion{C}{4} 1550~\AA \ line profile obtained by the two-dimensional
accretion disk wind model for an inclination angle of $25^\circ$.
The physical parameters here used are the same as in Figure~14~.}

\figcaption[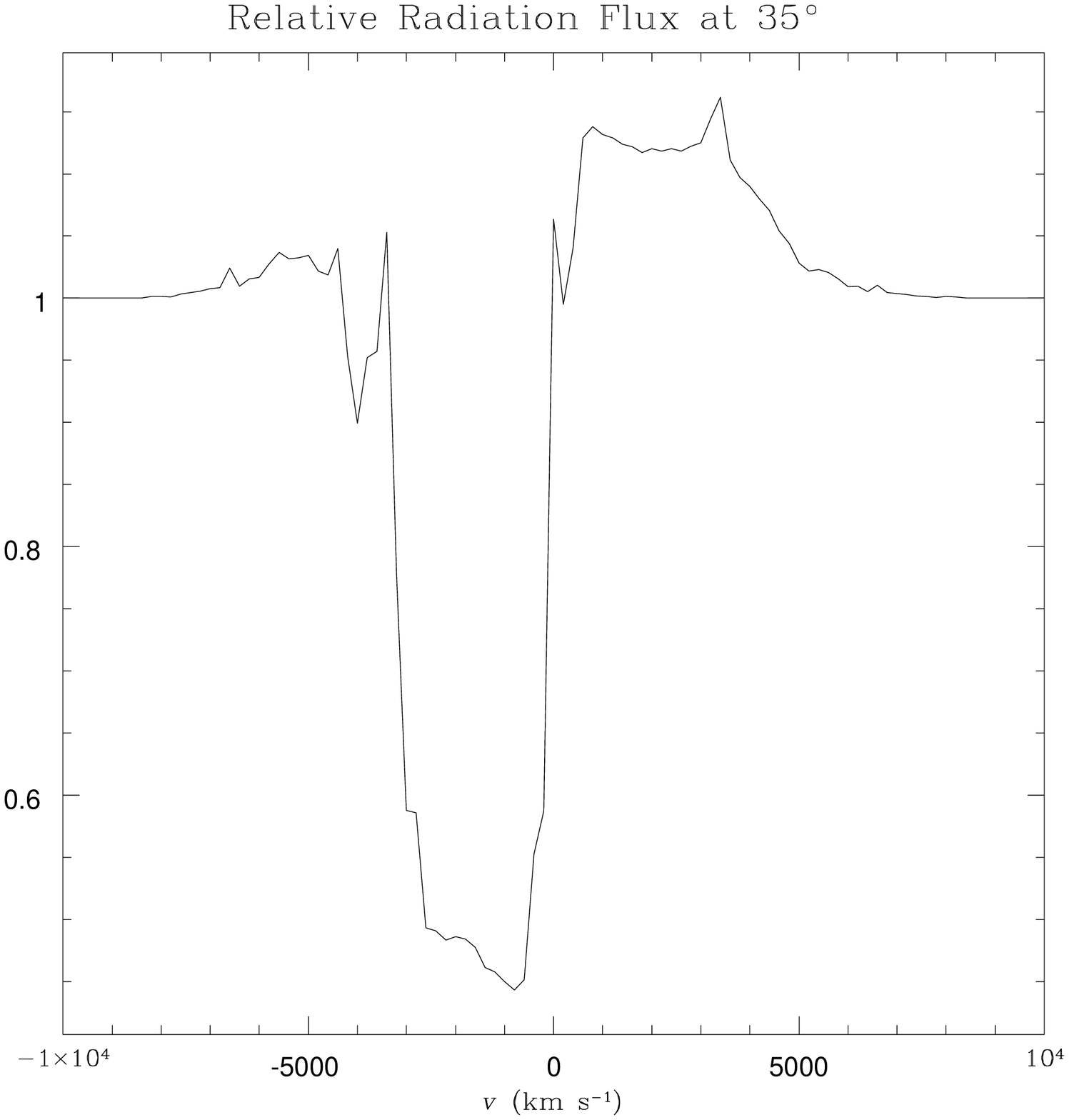]
{\ion{C}{4} 1550~\AA \ line profile obtained by the two-dimensional
accretion disk wind model for an inclination angle of $35^\circ$.
The physical parameters here used are the same as in Figure~14~.}

\figcaption[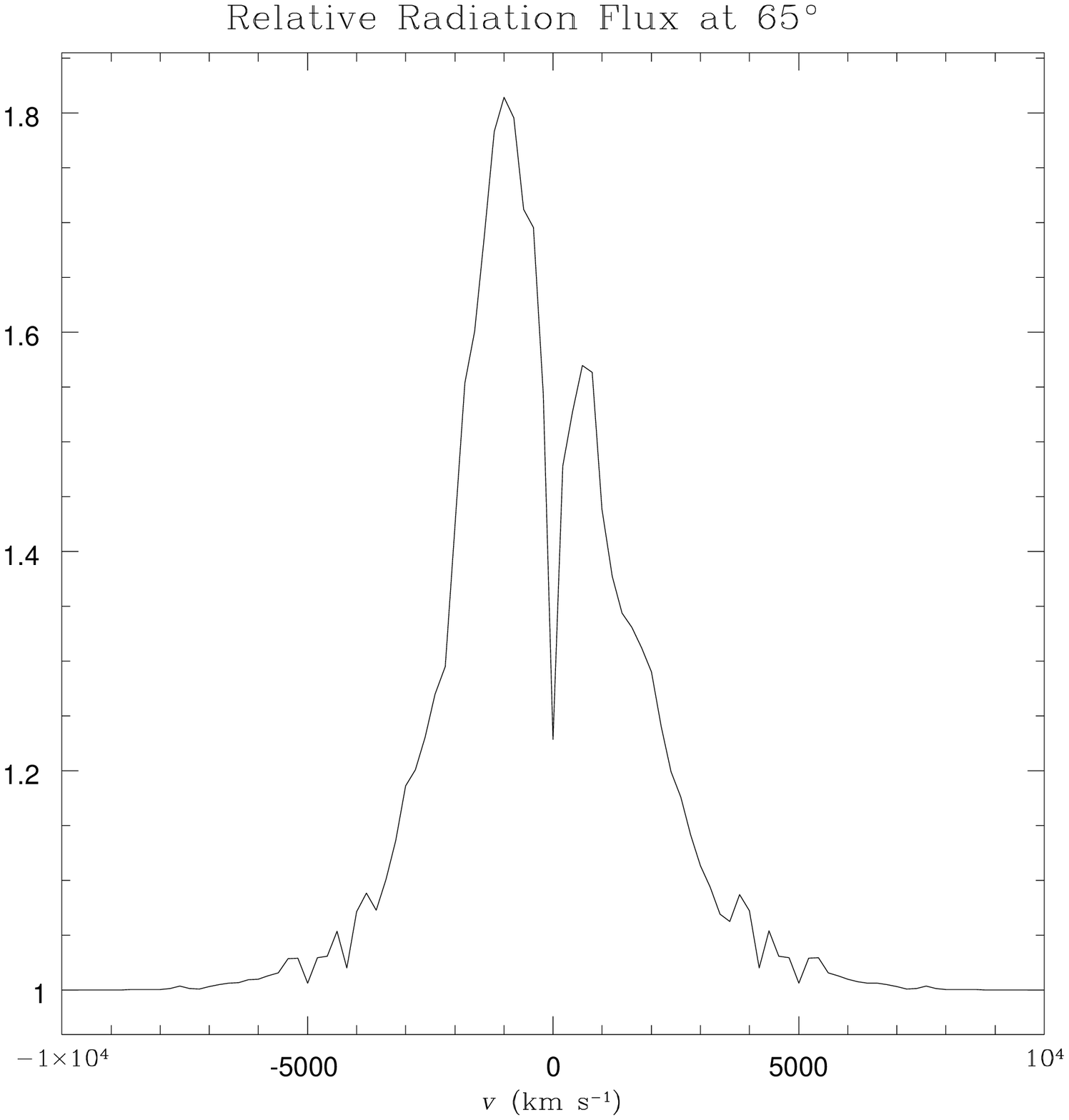]
{\ion{C}{4} 1550~\AA \ line profile obtained by the two-dimensional
accretion disk wind model for an inclination angle of $65^\circ$.
The physical parameters here used are the same as in Figure~14~.}

\figcaption[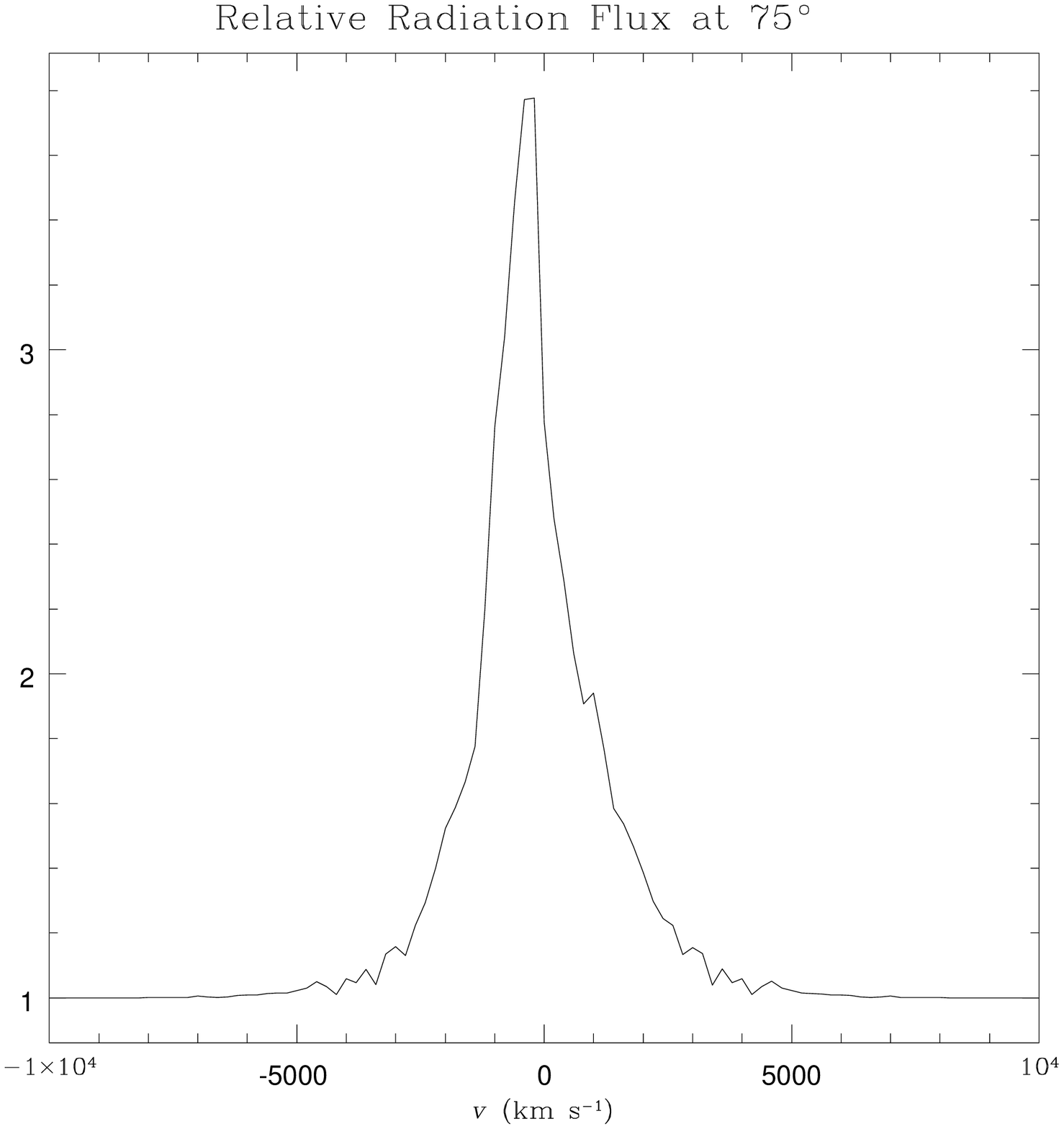]
{\ion{C}{4} 1550~\AA \ line profile obtained by the two-dimensional
accretion disk wind model for an inclination angle of $75^\circ$.
The physical parameters here used are the same as in Figure~14~.}

\figcaption[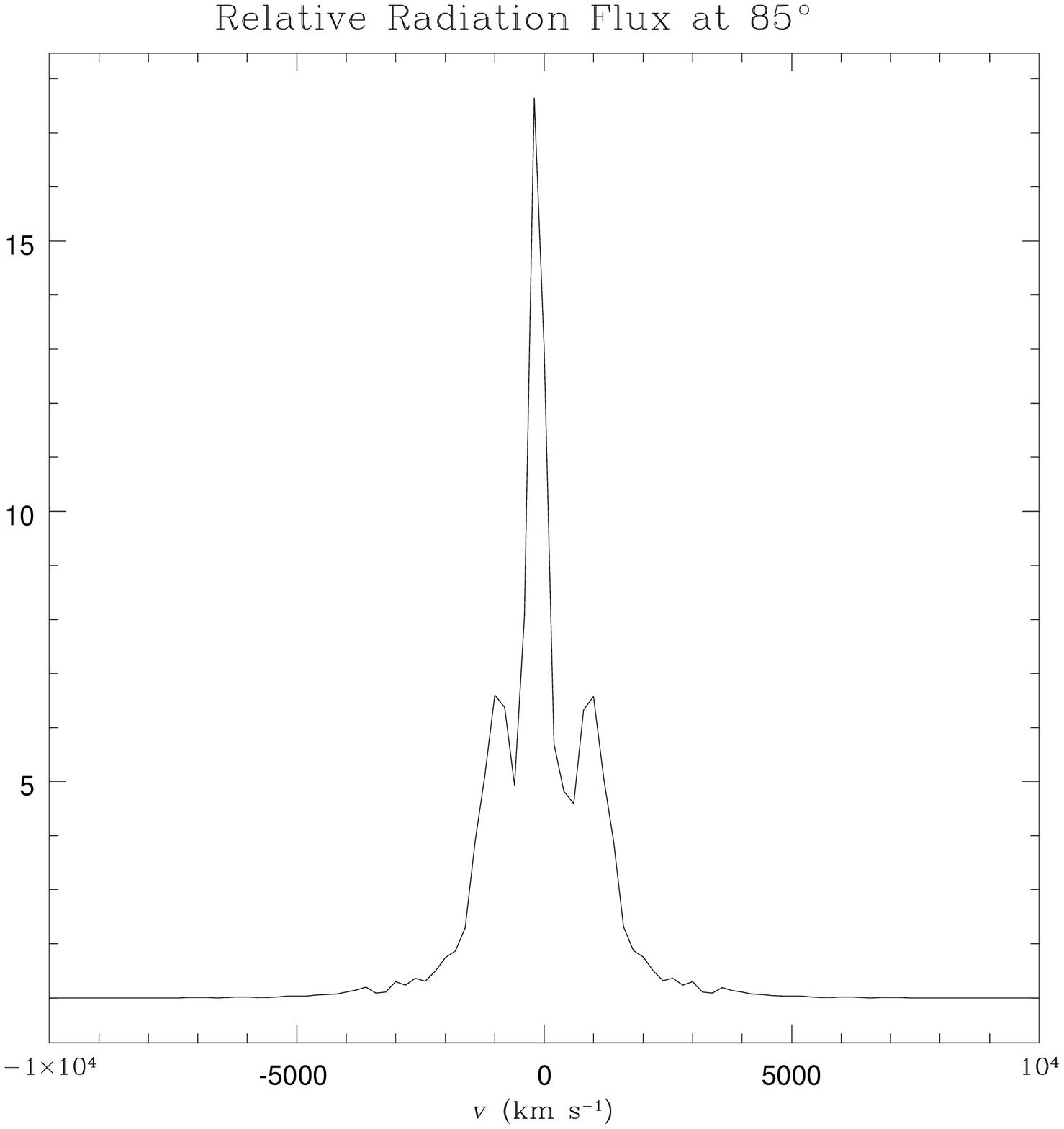]
{\ion{C}{4} 1550~\AA \ line profile obtained by the two-dimensional
accretion disk wind model for an inclination angle of $85^\circ$.
The physical parameters here used are the same as in Figure~14~.}

\clearpage

\begin{figure}
\figurenum{1}
\plotone{f1.eps}
\caption{}
\end{figure}

\begin{figure}
\figurenum{2}
\plotone{f2.eps}
\caption{}
\end{figure} 

\begin{figure}
\figurenum{3{\it a}}
\plotone{f3a.eps}
\caption{}
\end{figure} 

\begin{figure}
\figurenum{3{\it b}}
\plotone{f3b.eps}
\caption{}
\end{figure} 

\begin{figure}
\figurenum{3{\it c}}
\plotone{f3c.eps}
\caption{}
\end{figure} 

\begin{figure}
\figurenum{3{\it d}}
\plotone{f3d.eps}
\caption{}
\end{figure} 

\begin{figure}
\figurenum{3{\it e}}
\plotone{f3e.eps}
\caption{}
\end{figure} 

\begin{figure}
\figurenum{3{\it f}}
\plotone{f3f.eps}
\caption{}
\end{figure}

\begin{figure}
\figurenum{4}
\epsscale{0.5}
\plotone{f4.eps}
\caption{}
\end{figure}

\begin{figure}
\figurenum{5}
\epsscale{0.5}
\plotone{f5.eps}
\caption{}
\end{figure}

\begin{figure}
\figurenum{6}
\epsscale{0.5}
\plotone{f6.eps}
\caption{}
\end{figure}

\begin{figure}
\figurenum{7}
\epsscale{0.5}
\plotone{f7.eps}
\caption{}
\end{figure}

\begin{figure}
\figurenum{8}
\epsscale{0.5}
\plotone{f8.eps}
\caption{}
\end{figure}

\begin{figure}
\figurenum{9}
\epsscale{0.5}
\plotone{f9.eps}
\caption{}
\end{figure}

\begin{figure}
\epsscale{1.0}
\figurenum{10{\it a}}
\plotone{f10a.eps}
\caption{}
\end{figure}

\begin{figure}
\figurenum{10{\it b}}
\plotone{f10b.eps}
\caption{}
\end{figure}

\begin{figure}
\figurenum{11{\it a}}
\plotone{f11a.eps}
\caption{}
\end{figure}

\begin{figure}
\figurenum{11{\it b}}
\plotone{f11b.eps}
\caption{}
\end{figure}

\clearpage

\begin{figure}
\figurenum{12}
\epsscale{0.5}
\plotone{f12.eps}
\caption{}
\end{figure}

\begin{figure}
\figurenum{13{\it a}}
\epsscale{1.0}
\plotone{f13a.eps}
\caption{}
\end{figure}

\begin{figure}
\figurenum{13{\it b}}
\epsscale{1.0}
\plotone{f13b.eps}
\caption{}
\end{figure}

\begin{figure}
\figurenum{13{\it c}}
\epsscale{1.0}
\plotone{f13c.eps}
\caption{}
\end{figure}

\begin{figure}
\figurenum{14}
\epsscale{0.5}
\plotone{f14.eps}
\caption{}
\end{figure}

\begin{figure}
\figurenum{15}
\epsscale{1.0}
\plotone{f15.eps}
\caption{}
\end{figure}

\begin{figure}
\figurenum{16}
\epsscale{1.0}
\plotone{f16.eps}
\caption{}
\end{figure}

\begin{figure}
\figurenum{17}
\epsscale{1.0}
\plotone{f17.eps}
\caption{}
\end{figure}

\begin{figure}
\figurenum{18}
\epsscale{1.0}
\plotone{f18.eps}
\caption{}
\end{figure}

\begin{figure}
\figurenum{19}
\epsscale{1.0}
\plotone{f19.eps}
\caption{}
\end{figure}

\end{document}